\documentclass[journal]{IEEEtran}

\usepackage{cite}
\usepackage{amssymb}
\ifCLASSINFOpdf
   \usepackage[pdftex]{graphicx}
\else
   \usepackage[dvips]{graphicx}
\fi

\usepackage[cmex10]{amsmath}
\usepackage{mdwmath}
\usepackage[lined,boxed,commentsnumbered, ruled]{algorithm2e} 
\usepackage{amsthm} 

\usepackage{chemarrow}
\usepackage{extarrows}
\allowdisplaybreaks[4]
\usepackage{multirow}  
\usepackage{array} 
\usepackage{subfigure} 
\usepackage{stfloats}
\usepackage{color}
\usepackage[colorlinks,
            linkcolor=black,
            anchorcolor=black,
            urlcolor=black,
            citecolor=black]{hyperref}
\usepackage[hyphenbreaks]{breakurl}

\newcommand{\systemname}{LAENets}
\newcommand{\isacname}{ISAC}
\newcommand{\laename}{LAE}
\newcommand{\methodname}{DeDiff-VARARO}
\newcommand{\vddpgname}{VARARO}
\newcommand{\deisacname}{DeDiff-ISAC}
\begin{document}

\title{Vision-Aided ISAC in Low-Altitude Economy Networks via De-Diffused Visual Priors}
\author{Yulan Gao,\IEEEmembership{~Member, IEEE,}
Ziqiang Ye,\IEEEmembership{~}
Zhonghao Lyu,\IEEEmembership{~Member, IEEE,} 
Ming Xiao,\IEEEmembership{~Senior Member, IEEE,}
Yue Xiao,\IEEEmembership{~Member, IEEE,}
Ping Yang,\IEEEmembership{~Senior Member, IEEE,} 
and Agata Manolova,\IEEEmembership{~Member, IEEE} 
\thanks{Y. Gao, Z. Lyu and M. Xiao are with the Division of Information Science and Engineering, KTH Royal Institute of Technology, 100 44 Stockholm, Sweden (e-mail: yulang@kth.se, lzhon@kth.se, mingx@kth.se).}
\thanks{ Z. Ye, Y. Xiao and P. Yang are with the National Key Laboratory of Wireless Communications, University of Electronic Science and Technology of China (UESTC), Chengdu 611731, China (e-mail: yysxiaoyu@hotmail.com; xiaoyue@uestc.edu.cn, yang.ping@uestc.edu.cn).}
\thanks{A. Manolova is with the Faculty of Telecommunications, Technical University of Sofia, 1000 Sofia, Bulgaria (email: amanolova@tu-sofia.bg)}
}

\maketitle

\begin{abstract}
Emerging low-altitude economy networks (\systemname{}) require agile and privacy-preserving resource control under dynamic agent mobility and limited infrastructure support.
To meet these challenges, we propose a vision-aided integrated sensing and communication (\isacname{}) framework for UAV-assisted access systems, where onboard masked De-Diffusion models extract compact semantic tokens, including agent type, activity class, and heading orientation, while explicitly suppressing sensitive visual content.
These tokens are fused with mmWave radar measurements to construct a semantic risk heatmap reflecting motion density, occlusion, and scene complexity, which guides access technology selection and resource scheduling.
We formulate a multi-objective optimization problem to jointly maximize weighted energy and perception efficiency via radio access technology (RAT) assignment, power control, and beamforming, subject to agent-specific QoS constraints. To solve this, we develop De-Diffusion-driven vision-aided risk-aware resource optimization algorithm \methodname{}, a novel two-stage cross-modal control algorithm: the first stage reconstructs visual scenes from tokens via De-Diffusion model for semantic parsing, while the second stage employs a deep deterministic policy gradient (DDPG)-based policy to adapt RAT selection, power control, and beam assignment based on fused radar-visual states.
Simulation results show that \methodname{} consistently outperforms baselines in reward convergence, link robustness, and semantic fidelity, achieving within $4\%$ of the performance of a raw-image upper bound while preserving user privacy and scalability in dense environments.
\end{abstract}

\section{Introduction}
\subsection{Background}
\IEEEPARstart{T}{he evolving} landscape of wireless applications, ranging from autonomous aerial vehicles to immersive holographic and extended reality systems--is reshaping the design goals of future wireless networks. 
These applications demand not only high-throughput data exchange, but also real-time environmental awareness and intelligent response \cite{zhang20196g}. 
In current fifth-generation (5G) networks, the key functions of sensing, computing, and communication are treated separately, often lacking mutual reinforcement. 
While 5G excels in broadband connectivity and supports edge/cloud-based processing, it offers limited native support for environmental perception \cite{andrews2014will, mao2017survey}. 
This functional separation restricts the system's ability to adapt to complex, time-sensitive scenarios \cite{li2015deep}. 
A representative setting that highlights this limitation is the emerging low-altitude economy networks (\systemname{}), where airspace below 300 meters is increasingly utilized for logistics, aerial mobility, and infrastructure monitoring. 
\systemname{} environments are inherently dynamic and infrastructure-sparse, often rendering conventional ground-based networks insufficient. 
To meet the demands of such systems, it becomes essential to move beyond disjointed architectures and adopt an integrated sensing and communication (\isacname{}) paradigm--one that can enable airborne nodes to perceive, predict, and communicate efficiently in a coordinated fashion. 

\isacname{} has attracted growing attention from both academic and industry in recent years \cite{tan2021integrated,wymeersch2021integration, lyu2022joint}.
Early efforts primarily focused on spectrum sharing between radar and communication systems, aiming to improve spectral efficiency (SE) through coordinated but functionally separate designs \cite{gao2022integrated}. 
These approaches often relied on orthogonal allocation or interference mitigation strategies, which limited the overall system performance \cite{mahal2017spectral}.
Subsequent research introduced radar-centric schemes, where communication signals were embedded into radar waveforms. 
While promising in theory, these methods were constrained by the limited flexibility of radar signal structures, resulting in modest data rates. 
On the other hand, sensing-aided communication attempted to enhance wireless transmission by exploiting environmental awareness \cite{hassanien2016phase}. 
However, most existing designs still treat sensing and communication as loosely coupled modules, falling short of realizing the full potential of \isacname{}.
This calls for a deeper integration, where sensing and communication processes are jointly optimized and dynamically co-adaptive. 
In particular, \laename{} scenarios with their rapidly changing spatial structures and strict latency requirements demand an \isacname{} framework that can extract semantic context and guide physical-layer decisions in real time \cite{lu2024semantic, yang2025integrated}. 

In recent efforts to enhance millimeter-wave (mmWave) communication, researchers have explored the use of environmental awareness to guide beam selection. 
For example, Ref. \cite{xu20203d} proposed a camera-assisted strategy that integrates 3D geometry and material properties of surrounding structured settings. Such sensing-aided communication approaches often treat sensing as a supplementary module, rather than a core part of the transceiver design. 
Looking ahead to sixth-generation (6G) networks, integrated \isacname{} is expected to evolve from a peripheral enhancement to a fundamental design paradigm \cite{liu2022survey, xiao2024space}. 
The introduction of ultra-dense antenna arrays and the use of terahertz bands open the door to joint waveform design, where sensing and data transmission are performed simultaneously and adaptively \cite{hua2023optimal, wild2021joint}.
These capabilities promise not only greater spectral efficiency (SE), but also the ability to dynamically perceive and respond to complex spatial environments in real time.

\laename{} environments present a unique set of challenges and opportunities for \isacname{} \cite{jiang2025integrated, cheng2025networked}.
In dense urban settings of infrastructure-sparse regions, conventional ground-based base stations (BSs) often suffer from limited line-of-sight (LoS) and constrained perception capabilities \cite{tang2025cooperative}. 
The presence of dynamic aerial agents, unpredictable obstacles, and rapidly changing trajectories further complicates sensing and beam alignment. 
These factors call for elevated, mobile platforms equipped with both communication and perception modules. 
UAV-mounted BSs (UAV-BSs) offer a compelling solution to these issues.
Operating at altitude, UAV-BSs can flexibly reposition and maintain LoS with mobile users while leveraging onboard visual sensors and mmWave radar to perceive the environment form a bird's-eye view. 
This elevated perspective allows them not only to extend communication coverage, but also to construct a semantic understanding of the surrounding space, capturing regions of high mobility density, visual occlusion, or signal obstruction. 

\subsection{Motivations and Contributions}
While some recent studies have introduced visual information into \isacname{} frameworks mainly to improve localization or assist beam prediction, the use of high-level semantic features from visual scenes to guide communication resource scheduling remains underexplored \cite{feng2025networked, ye2024integrated}.  
In particular, the potential to integrate structured visual context into user prioritization, radio access technology optimization, or access adaptation has not been systematically addressed.
Moreover, most existing approaches overlook the broader spectrum of privacy risks that arise when raw visual data is transmitted or processed centrally. 
These risks extend beyond user identity to include the leakage of location-sensitive features, recognizable landmarks, mobility patterns, and structural cues that may enable unauthorized scene reconstruction or spatial inference. 
This is especially critical for UAV-based ISAC systems operating in public or strategically sensitive airspaces, where perceptual data may unintentionally expose protected physical or operational information.
Additionally, the limited transmission power and energy budget of UAV platforms, coupled with the high bandwidth requirements of raw image transmission, especially under mmWave communication constraints, make such centralized visual data exchange impractical. 
These challenges call for lightweight, semantic abstractions of visual input that retain behavioral semantics while eliminating identifiable environmental cues and reducing overhead.

To address the above limitations, this paper proposes a vision-aided \isacname{} framework specifically designed for UAV-assisted communication in \laename{} environments. Typical LAE applications, such as drone-enabled emergency response, smart city logistics, and large-scale aerial monitoring, often operate in highly dynamic and cluttered environments. 
These scenarios are characterized by dense infrastructure, occluded urban topologies, fast-changing user mobility, and the lack of fixed sensing or communication infrastructure, all of which demand agile and semantically-aware communication strategies.
The proposed system leverages semantic cues from onboard visual sensors to inform communication decisions, while preserving operational privacy and minimizing transmission overhead.

The main contributions are summarized as follows:

\begin{itemize}
\item{\bf\em{Vision-Aided \isacname{} with De-Diffusion:}} We propose a vision-aided \isacname{} framework in which UAV-mounted cameras capture real-time visual data from low-altitude environments. 
To reduce transmission overhead and avoid scene-level reconstruction risks, a masked de-diffusion model is deployed onboard the UAV to extract high-level semantic tokens--such as motion type, heading direction, and activity class.  
These compact and privacy-aware tokens are then transmitted to the cloud server, where they assist downstream \isacname{} tasks such as beamforming and RAT selection without requiring raw image transmission.
The extracted tokens are stripped of spatially identifiable textures and structural cues, preventing the transmission of sensitive scene content while maintaining \isacname{} utility. 

\item{\bf\em{Vision-Assisted Risk Map for Scheduling Guidance:}} A visual-semantic risk map is constructed at the cloud server by fusing high-level semantic tokens extracted via de-diffusion from UAV-acquired imagery with mmWave radar measurements. 
These tokens capture behavioral patterns such as motion density, heading alignment, and activity class, while mmWave data provides quantitative estimates of relative velocity, spatial proximity, and potential occlusions. 
The fused representation, obtained by parsing reconstructed images via YOLOv11, enables the construction of a dynamic risk map that reflects both scene-level complexity and physical-layer interaction intensity.
This risk map serves as a scheduling prior to guide user prioritization and adaptive RAT selections within the ISAC framework\footnote{Note that the construction of the risk map is based solely on abstracted semantic tokens and physical-layer measurements, neither of which contain raw visual data or spatially reconstructable features. This design ensures that while behavioral complexity and interaction risk can be quantified, the underlying scene content remains obscured.}.
\item{\bf\em{De-Diffusion-Driven Vision-Aided Risk-Aware Resource Optimization Algorithm (\methodname{}):}}  
We formulate the \isacname{} resource control problem as a continuous-space optimization task and introduce a DDPG-based agent that jointly optimizes energy efficiency, link stability, and visual-semantic risk mitigation.  
Distinct from existing works, our agent observes a cross-modal state space that fuses mmWave sensing data with de-diffused semantic tokens, including motion type, heading, and activity class.  
These features enable the agent to anticipate environmental complexity and make fine-grained decisions on RAT selection and power allocation.  
A novel visual-risk-aware reward function further guides the learning agent to prioritize users in congested, occluded, or conflict-prone zones, promoting scheduling robustness under \systemname{} dynamics.
\end{itemize}

Our work is inspired by recent vision-assisted \isacname{} studies \cite{xu2022computer, alrabeiah2020millimeter}, which leverage camera data to improve beam alignment or blockage prediction. 
While these efforts highlight the potential of visual inputs for physical-layer enhancement, they primarily operate at the raw image or feature level and do not establish a semantic representation pipeline that supports cross-layer decision-making. 
Moreover, considerations such as privacy preservation, transmission overhead, and dynamic scheduling have not been systematically addressed, particularly in highly dynamic, infrastructure-limited settings like those found in \systemname{}.
In contrast, our approach introduces a semantic-token-driven \isacname{} framework that abstracts visual content through a masked de-diffusion process and integrates the resulting representation with mmWave radar feedback to support cloud-side scheduling. This allows environmental semantics to inform not just perception, but access decisions as well.

\subsection{Outline of Paper}
The rest of this paper is structured as follows. In Section II, we summarize the related work. 
Section III specifies the system overview, De-Diffusion-based visual token extraction, mmWave radar-based agent localization, communication model for mmWave and LTE, and RAT-aware risk-informed scheduling logic. 
The system state representation and problem formulation are presented in Section IV. 
Section V describes in detail how we solve the formulated optimization problem by \methodname{}.
Simulation results are shown in Section VI and we conclude in Section VII.

\section{Related Work}
\subsection{Vision-Aided ISAC and Context-Aware Scheduling}

Recent advances in ISAC have shown growing interest in leveraging visual context to enhance physical-layer adaptation, particularly in highly dynamic and infrastructure-sparse environments such as LAENets. 
Existing studies have primarily utilized environmental features (e.g., depth, geometry, material type) from RGB or LiDAR sensors to guide beam prediction or blockage detection in mmWave communications~\cite{xu20203d,xu2022computer, yuan2025ground}. However, these methods often rely on raw image transmission or heavy visual feature extraction pipelines, raising both scalability and privacy concerns.
Moreover, while some ISAC works incorporate perception feedback to improve beamforming or handover, they typically lack semantic abstraction and treat visual signals as auxiliary sources~\cite{alrabeiah2020millimeter,liu2022survey}. In contrast, our work introduces a cross-modal semantic integration strategy, in which structured visual tokens derived via masked De-Diffusion are fused with radar sensing outputs to inform access control and resource scheduling. This enables UAVs not only to perceive environmental complexity but also to make semantically aware decisions in real-time.

\subsection{RAT Selection and Risk-Aware Access Control}
Multi-RAT architectures especially those combining LTE and mmWave have become essential in adapting to heterogeneous link conditions and mobility profiles~\cite{liu2020joint, zhou2025sign}. 
Prior efforts have focused on utility-aware access control and beam management using reinforcement learning or heuristic methods~\cite{liu2014multiobjective,annaswamy2023adaptive,zhao2025temporal}. These approaches, however, often assume full observability of agent state or ideal link-level measurements, overlooking latent behavior patterns such as group mobility, occlusion-induced degradation, or privacy-relevant positioning.

To address this gap, we incorporate semantic profiles, comprising agent type, activity class, and heading estimate, into the access control loop, constructing a risk-aware visual heatmap that informs both RAT selection and prioritization logic. 
Our work advances prior art by integrating perception uncertainty directly into the scheduling policy, rather than treating it as a posterior metric.
\subsection{Privacy-Preserving Visual Modeling via De-Diffusion}
Traditional vision-based systems often transmit raw or partially masked images to the edge/cloud, risking exposure of sensitive information such as identifiable landmarks or user trajectory patterns. To mitigate this, recent studies have explored privacy-preserving learning through adversarial masking, differential privacy, or federated frameworks~\cite{wang2020learning,zhu2023pushing, yang2021privacy}.
Yet, few works have addressed the unique trade-off between visual abstraction and ISAC utility in airborne networks.

Our proposed approach builds on masked De-Diffusion modeling~\cite{wei2024diffusion} to generate structured text tokens that describe semantic intent (e.g., ``cyclist moving east'') while suppressing spatial and texture-level cues. These tokens are further reconstructed into synthetic images via pretrained diffusion models (e.g., StableXL), enabling downstream semantic parsing with no raw visual exposure. This paradigm aligns with the growing trend in cross-modal privacy-preserving learning, but is tailored for the real-time, energy-constrained, and perception-dependent nature of LAENets.

\section{System Model}
\subsection{System Overview}
\begin{figure*}[!t]
\centering
\includegraphics[width=1\textwidth]{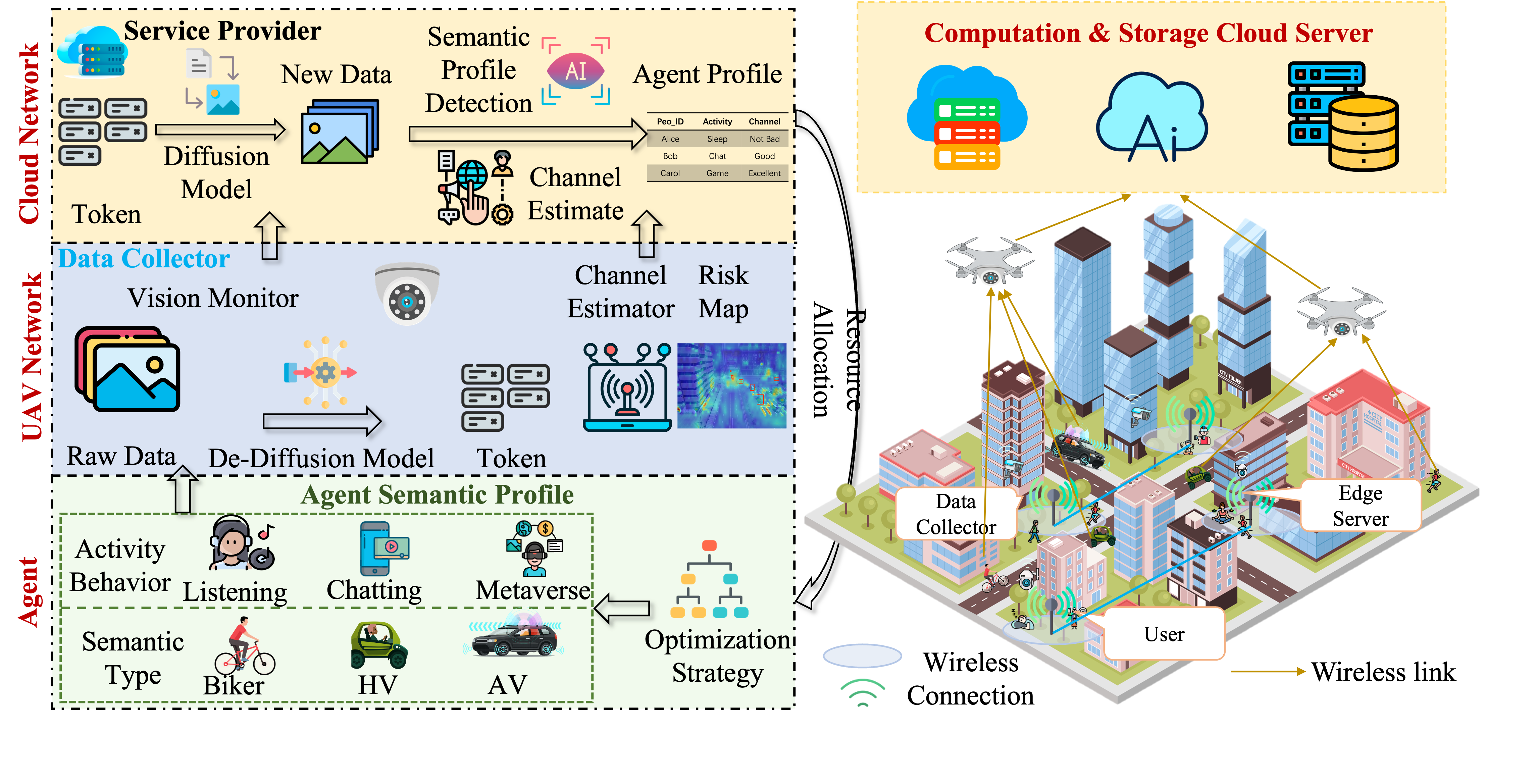}
\caption{System architecture of the proposed vision-aided ISAC framework in \systemname{}. The architecture is composed of three tiers: (i) the ground layer, where agents of different semantic types (e.g., bikers, human vehicles (HV), autonomous vehicles (AV)) are classified by their activity behavior (e.g., listening, chatting, metaverse participation); (ii) the UAV network layer, where onboard cameras and mmWave radar perform cross-modal sensing, and a masked de-diffusion model extracts privacy-preserving semantic tokens from visual data; and (iii) the cloud network layer, where semantic tokens are uploaded for image reconstruction, semantic profile detection, and channel estimation. A risk-aware heatmap is constructed based on YOLOv11 parsing and radar sensing to guide resource allocation. The output agent profile is used to generate optimization strategies for RAT selection, beam assignment, and power control. The entire system operates in a closed loop to support dynamic access control and semantic-level ISAC in infrastructure-sparse environments.}
\label{fig:1}
\end{figure*}

We consider a UAV-assisted \isacname{} architecture deployed in a \laename{} environment.
The UAV is equipped with three components: a visual sensing unit (i.e., an onboard RGB camera), a mmWave radar module, and a communication transceiver supporting multiple RATs, such as LTE and mmWave bands. 
The UAV hovers or patrols at a moderate altitude (e.g., $100-150 \mbox{m}$), providing simultaneous perception and communication coverage over a representative segment of a smart city characterized by dense buildings, streets, and intersections, which introduce frequent occlusions and severe multipath effects, as illustrated in Fig. \ref{fig:1}. 

Consider $N$ ground agents located within the UAV’s coverage area, and let the agent set be denoted as $\mathcal{N} = \{1, 2, \dots, N\}$.
Each agent $n \in \mathcal{N}$ is associated with a semantic profile $\mathrm{sf}_n = (\mathrm{sem}_n,\ \mathrm{act}_n)$, where  $\mathrm{sem}_n \in \mathcal{C}$ denotes the agent’s semantic type (e.g., pedestrian, vehicle, cyclist), and
$\mathrm{act}_n \in \mathcal{G}$ represents the current activity class (e.g., moving, turning, stopping).
Here, $\mathcal{C} = \{1, 2, \ldots, C\}$ and $\mathcal{G} = \{1, 2, \ldots, G\}$ denote the finite sets of possible semantic types and activity classes, respectively.
The joint semantic space is defined as the Cartesian product:
$\mathcal{S} = \mathcal{C} \times \mathcal{G},$
which enumerates all possible semantic profiles.
Subsequently, the semantic profile collection for all agents is denoted as: $\mathbf{sf} = \left\{ \mathrm{sf}_n = (\mathrm{sem}_n, \mathrm{act}_n) \ \big| \ n \in \mathcal{N} \right\}.$
The key notation definitions are summarized in Table \ref{table:1}.

\begin{table}[!t]
\centering
\caption{List of Notations}
\resizebox{1\linewidth}{!}{
\begin{tabular}{|c|p{2.55cm}|c|p{2.55cm}|}
  \hline
  ${\cal N}$ & set of agents & ${\cal G}$ & set of activity\\
  \hline
  ${\mathcal C}$ &set of semantic type & $\cal S$ & set of semantic profile\\
\hline
$\mathrm{DeDiff(\cdot)}$ &\multicolumn{3}{p{6.5cm}|}{the process of De-Diffusion model}\\
\hline
$I_t^n, \hat{I}_t^n$ &\multicolumn{3}{p{6.5cm}|}{raw image, reconstructed synthetic image}\\
\hline 
$\mathbf{z}_{\text{text},t}^n$ &\multicolumn{3}{p{6.5cm}|}{structured textual token by De-Diffusion model}\\
\hline
$\mathbf{z}_{\text{vis},t}^n$ &\multicolumn{3}{p{6.5cm}|}{recognized agent semantic profile from $\hat{I}_t^n$}\\
\hline
$\theta_{\text{vis},t}^n$ &\multicolumn{3}{p{6.5cm}|}{heading orientation }\\
\hline
$d_n,v_n,\psi_n$&\multicolumn{3}{p{6.5cm}|}{estimated distance, velocity and angle of agent $n$}\\
\hline
$B$ &\multicolumn{3}{p{6.5cm}|}{transmitted signal sweep frequency}\\
\hline
$T$ &\multicolumn{3}{p{6.5cm}|}{ signal frequency raising cycle}\\
\hline
\multirow{2}*{$f_0, f_n$} &\multicolumn{3}{p{6.5cm}|}{center frequency of transmitted signal and its corresponding intermediate frequency}\\
\hline
{$\gamma_n$}&\multicolumn{3}{p{6.5cm}|}{SINR for agent $n$}\\
\hline
$L$&\multicolumn{3}{p{6.5cm}|}{total number of symbols in a time slot}\\
\hline
$L_p$&\multicolumn{3}{p{6.5cm}|}{the number of pilot symbols}\\
\hline
$\pmb{w}$&\multicolumn{3}{p{6.5cm}|}{precoding matrix}\\
\hline
$\pmb{H}_n$&\multicolumn{3}{p{6.5cm}|}{normalized narrowband millimeter wave channel}\\
\hline
$\mathcal{L}$&\multicolumn{3}{p{6.5cm}|}{the number of scattering paths}\\
\hline
$\pmb{a}(\phi_l)$&\multicolumn{3}{p{6.5cm}|}{steering vector of the $l$-th path transmitting side}\\
\hline
$p_n$&\multicolumn{3}{p{6.5cm}|}{the transmit power for agent $n$}\\
\hline
{$R_n$}&\multicolumn{3}{p{6.5cm}|}{the achieved data rate between agent $n$ and UAV}\\
\hline
$\lambda$&\multicolumn{3}{p{6.5cm}|}{the wavelength of carrier wave}\\
\hline
$s_{\text{rcs}}$&\multicolumn{3}{p{6.5cm}|}{radar cross-sectional area}\\
\hline
$T_s$&\multicolumn{3}{p{6.5cm}|}{symbol time interval}\\
\hline
$B_{\text{rms}}$&\multicolumn{3}{p{6.5cm}|}{rms bandwidth}\\
\hline
$A_{s}$&\multicolumn{3}{p{6.5cm}|}{main path channel strength}\\
\hline
$\sigma_1,\sigma_2$&\multicolumn{3}{p{6.5cm}|}{the variance of the perception channel}\\
\hline
$\lambda^{EE},\lambda^{PE}$&\multicolumn{3}{p{6.5cm}|}{balance parameter for utility function}\\
\hline
$p^{\max}$&\multicolumn{3}{p{6.5cm}|}{maximum power limitation}\\
\hline
\end{tabular}}
\label{table:1}
\end{table}

\subsection{De-Diffusion-Based Visual Token Extraction for \isacname{} Systems}

\begin{figure}[!t]
\centering
\includegraphics[width=0.45\textwidth]{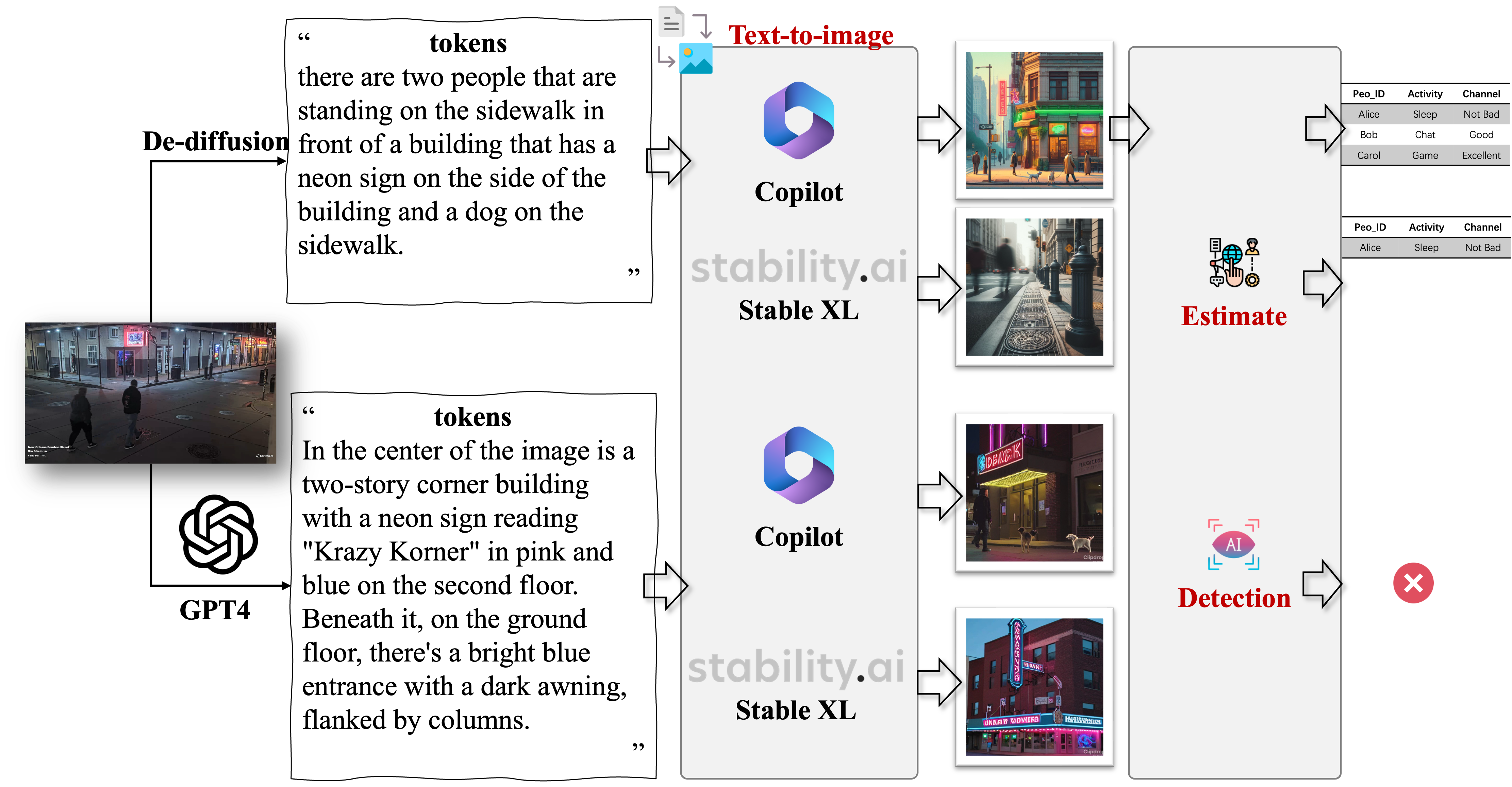}
\caption{De-Diffusion model and GPT4 Assisted multi-stage visual data processing for semantic profile classification in \systemname{}, respectively.}
\label{fig:5}
\end{figure}

To enable accurate semantic-level perception and privacy-preserving sensing in intelligent \systemname{}, we adopt a vision-aided \isacname{} framework that integrates mmWave radar sensing with de-diffused visual priors.
At each time step $t$, the onboard camera captures an image $I_t^n$ for agent $n$.
Instead of directly uploading raw visual data, we leverage a masked De-Diffusion model \cite{wei2024diffusion} to transform $I_t^n$ into {\em structured textual tokens} $\mathbf{z}_{\text{text},t}^n$ that describe only coarse semantic attributes while omitting sensitive visual cues, an example of its operation process is illustrated in Fig. \ref{fig:5}. 
The model explicitly removes privacy-sensitive regions, such as identifiable background elements or personal belongings from the visual input, retaining only task-relevant features. 
The resulting representation is defined as:  
\begin{equation}
\mathbf{z}_{\text{text},t}^n = \mathrm{DeDiff}(I_t^n).
\end{equation}
The tokens are transmitted to the edge server and reconstructed through a reverse diffusion model \cite{croitoru2023diffusion} to reconstruct a synthetic embedding $\hat{I}_t^n$, which is further fed into {\bf YOLOv11} \cite{yolo} and {\bf SlowFast} models \cite{fan2020pyslowfast} to recognize agent semantic profile $\mathbf{z}_{\text{vis},t}^n \in \mathcal{S}$.
Beyond agent semantic profile classification, we further exploit the semantic profile $\mathbf{z}_{\text{vis},t}^n$ to enhance the radar-based sensing pipeline. 
Specifically, we distinguish between the agent’s semantic type (e.g., vehicle, pedestrian, cyclist) and its current activity state (e.g., stopping, walking, crossing), where the former defines its physical profile and motion capability, and the latter captures its instantaneous behavioral pattern. Both are included in the semantic token to support risk estimation and access prioritization.

Here, in addition to semantic profile classification, the de-diffusion-derived semantic profile $\mathrm{sf}_t^n$ includes the {\em heading estimate} $\theta_{\text{vis},t}^n\in[0, 2\pi)$, representing the forward-facing direction (yaw) of the agent in the global frame. 
This is extracted by applying semantic pose estimation to the visual reconstruction $\hat{I}_t^n$, and provides directional cues even for static agents. 
Importantly, $\theta_{\text{vis},t}^n$ is not equivalent to the motion direction implied by the radar-estimated velocity vector 
$v_t^n$. 
The heading estimate $\theta_{\text{vis},t}^n$ is used as a prior for mmWave beam alignment from the codebook $\mathcal B$: 
\begin{align}\label{eq:3}
\hat{b}_t^n=\arg\max_{b\in{\mathcal B}} \cos(\theta_b-\theta_{\text{vis},t}^n).
\end{align}
\subsection{mmWave Radar-based Agent Localization}
Beyond its role in assisting communication beamforming and risk-aware scheduling, the integrated mmWave radar module on the UAV also performs direct user localization as part of the ISAC framework. 
This capability is critical for maintaining accurate spatial awareness of all users in dynamic, infrastructure-free environments.

The radar transceiver adopts a frequency-modulated continuous wave (FMCW) waveform for ranging and velocity estimation. 
Each uplink radar sweep yields a reflected signal containing three types of information: range $d_n$, radial velocity $v_n$, and angular displacement $\psi_n$ for each detected user $n \in \mathcal{N}$.

\begin{figure}[!t]
\centering
\includegraphics[width=0.4\textwidth]{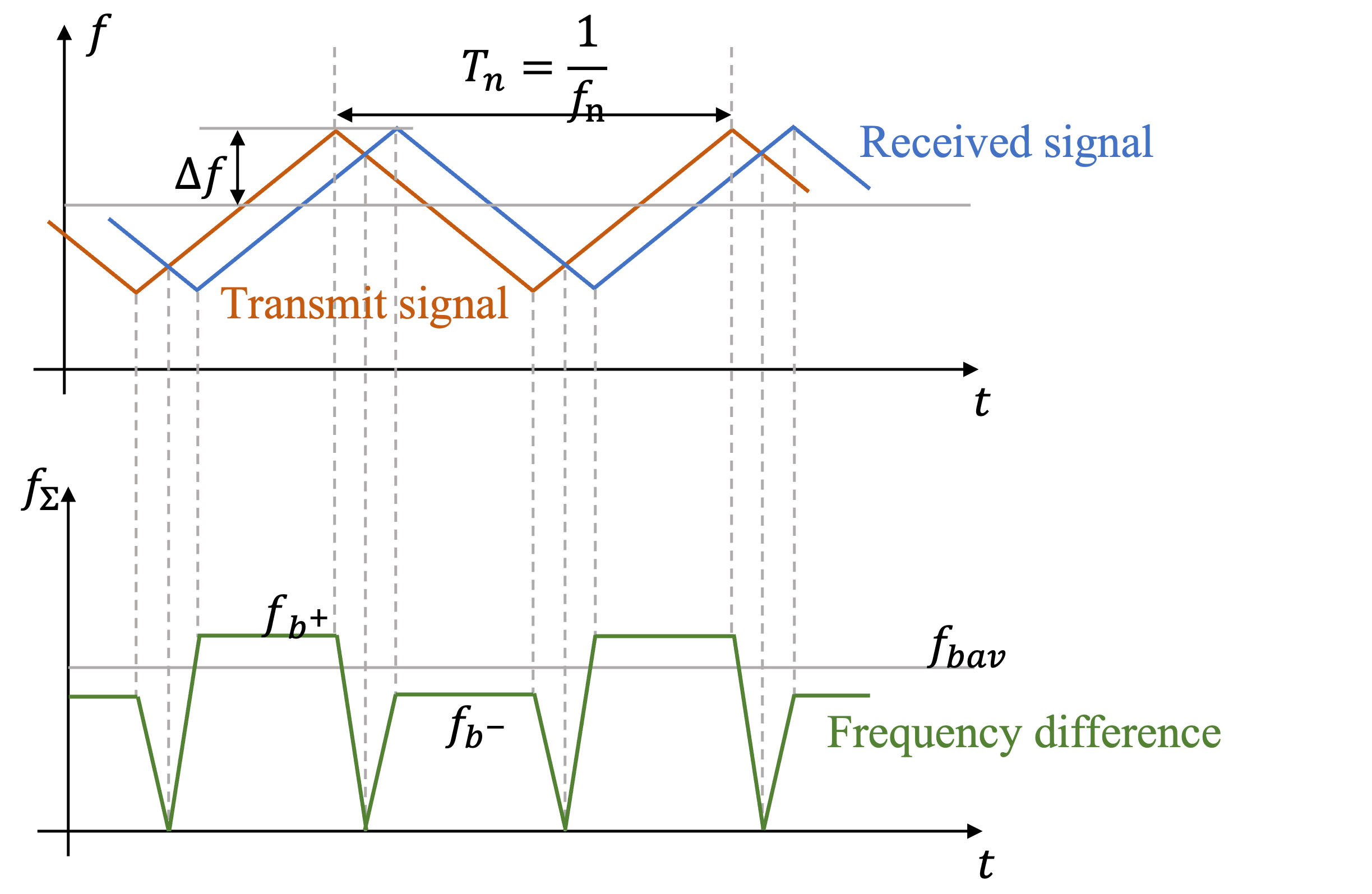}
\caption{Distance estimate.}
\label{fig:2}
\end{figure}

{\em Range Estimation:}
The beat frequency $f_n$ of the radar echo is linearly proportional to the user’s distance $d_n$, as depicted in Fig. \ref{fig:2}:
\begin{align}
d_n = \frac{c T f_n}{2 B},
\end{align}
where $c$ is the speed of light, $T$ is the chirp duration, and $B$ is the sweep bandwidth.

{\em Velocity Estimation:}
The radial velocity $v_n$ is estimated from the Doppler frequency shift $\omega_n$ across successive chirps:
\begin{align}
v_n = \frac{\lambda \omega_n}{4\pi T_s},
\end{align}
where $\lambda$ is the carrier wavelength and $T_s$ is the pulse repetition interval.

{\em Angle Estimation:}
The angular position of the user relative to the UAV’s antenna array is inferred from phase differences across antenna elements:
\begin{align}
\psi_n = \sin^{-1} \left( \frac{\lambda \omega_n}{2\pi d} \right),
\end{align}
where $d$ is the inter-element antenna spacing.

Together, the triplet $(d_n, v_n, \psi_n)$ forms the radar-based spatial state of agent $n$, enabling the construction of a 2D or 3D agent map. 
This localization output not only supports downlink beam steering and channel selection, but also acts as a standalone perception layer for trajectory tracking, obstacle avoidance, and predictive scheduling.

The spatial estimates are periodically fused with visual semantic profile $\mathrm{sf}^n$ to enhance robustness, especially under NLoS conditions or occlusions. 
This fusion is further leveraged in the scheduling logic described in Section~\ref{sec:RAT_control}.
\subsection{Communication Model for mmWave and LTE}

In the proposed \isacname{} system, each agent $n \in \mathcal{N}$ dynamically selects one of two available RATs: high-frequency mmWave or sub-6 GHz LTE. 
Due to their distinct propagation characteristics and physical-layer implementations, we adopt different channel and SINR models for each RAT.

{\em $\bullet$ mmWave Channel and SINR Model:}
The mmWave channel between the UAV and agent $n$ is modeled as a sparse geometric channel with $L$ resolvable paths:
\begin{align}
\mathbf{H}_n^{\text{mm}} = \sqrt{\frac{M}{L}} \sum_{\ell=1}^{L} \alpha_\ell \mathbf{a}(\varphi_\ell),
\end{align}
where $\alpha_\ell$ is the complex gain of the $\ell$-th path, $\varphi_\ell$ is its angle of departure (AoD) of the $\ell\mbox{th}$ path which is generally considered to be uniformly distributed within $[-\frac{\pi}{2},\frac{\pi}{2}]$, $M$ is the number of transmit antennas, and $\mathbf{a}(\varphi_\ell)$ is the uniform linear array (ULA) steering vector of the $\ell\mbox{th}$ path transmitting side: 
\begin{align}
    \mathbf{a}(\phi_l)=\sqrt{\frac{1}{M}}\left[1,e^{j\frac{2\pi}{\lambda}d\sin(\phi_l)},\ldots,e^{j\frac{2\pi}{\lambda}d(M-1)\sin(\phi_l)}\right]^T,
\end{align}
Given the transmit beamforming vector $\mathbf{w}_n \in \mathbb{C}^{M \times 1}$ assigned to agent $n$, the received SINR is:
\begin{align}
\gamma_n^{\text{mm}} = \frac{|\mathbf{H}_n^{\text{mm}H} \mathbf{w}_n|^2}{\sum_{i \neq n} |\mathbf{H}_n^{\text{mm}H} \mathbf{w}_i|^2 + \sigma^2},
\end{align}
where $\sigma^2$ denotes the noise power. The corresponding achievable rate is:
\begin{align}
R_n^{\text{mm}} = B_n \left(1 - \frac{L_p}{L} \right) \log_2(1 + \gamma_n^{\text{mm}}),
\end{align}
with $B_n$ denoting bandwidth, $L$ the frame length, and $L_p$ the number of pilot symbols per frame.

{\em $\bullet$ LTE Channel and SINR Model: }
For LTE access, we assume a flat fading channel modeled as:
\begin{align}
H_n^{\text{LTE}} \sim \mathcal{CN}(0, \sigma_h^2),
\end{align}
where $\sigma_h^2$ is the average channel gain depending on distance-based path loss.
The LTE SINR for agent $n$ is modeled as:
\begin{align}
\gamma_n^{\text{LTE}} = \frac{P_n |H_n^{\text{LTE}}|^2}{\sigma^2},
\end{align}
assuming orthogonal resource allocation and negligible inter-user interference. 
The corresponding achievable rate is:
\begin{align}
R_n^{\text{LTE}} = B_n \log_2(1 + \gamma_n^{\text{LTE}}).
\end{align}

{\em $\bullet$ RAT-Aware Access Decision:}
At each scheduling interval, agent $n$ selects its RAT via a binary variable $x_n \in \{0, 1\}$, where $x_n = 1$ denotes mmWave and $x_n = 0$ denotes LTE. 
The overall data rate is given by:
\begin{align}
R_n = x_n R_n^{\text{mm}} + (1 - x_n) R_n^{\text{LTE}}.
\end{align}

\subsection{RAT-Aware Risk-Informed Scheduling Logic}
\label{sec:RAT_control}
To enable adaptive and situation-aware connectivity, the system integrates semantic perception and channel state information into a unified scheduling logic. 
The goal is to assign each agent $n \in \mathcal{N}$ an appropriate access technology (mmWave or LTE) and resource configuration, based on real-time visual and radio conditions.

{\em Priority Estimation:}
Each agent is first assigned a scheduling priority score $\rho_n$ that reflects its contextual urgency and link suitability:
\begin{align}
\rho_n = \phi\left(\mathbf{z}_{\text{vis}}^n, \theta_{\text{vis},t}^n, \mathbf{H}_{\text{vis}}(x_n, y_n), \gamma_n^{\text{mm}}, \text{LoS}_n \right),
\end{align}
where $\mathbf{z}_{\text{vis}}^n$ and $\theta_{\text{vis},t}^n$ respectively represent semantic profile and heading, $\mathbf{H}_{\text{vis}}(x_n, y_n)$ is visual risk score at the agent's location, and $\text{LoS}_n$ is LoS availability (binary).
The mapping function $\phi(\cdot)$ can be rule-based or learned (e.g., via neural network), and encodes policies such as i) Prioritize users in high-risk or crowded zones;
ii) Downgrade users under occlusion or low mmWave SINR;
iii) Favor users with directional consistency between visual heading and radar angle.

{\em RAT Selection:}
Based on the priority score $\rho_n$, each agent selects its RAT using a soft-thresholding mechanism:
\begin{align}
x_n =
\begin{cases}
1, & \text{if } \rho_n \geq \delta_{\text{mm}} \& \text{LoS}_n = 1, \\
0, & \text{otherwise},
\end{cases}
\end{align}
where $\delta_{\text{mm}}$ is a tunable scheduling threshold. This ensures that mmWave access is granted to users who both require high-resolution connectivity and possess reliable visual-radar conditions.

{\em Resource Awareness: }
The final RAT allocation $\mathbf{x} = \{x_n\}_{n=1}^N$ is subject to resource constraints:
\begin{align}\label{eq:16}
\sum_{n=1}^N x_n \leq N_{\text{mm}}^{\max}, \quad \sum_{n=1}^N (1 - x_n) \leq N_{\text{LTE}}^{\max},
\end{align}
where $N_{\text{mm}}^{\max}$ and $N_{\text{LTE}}^{\max}$ denote the available access capacity for each RAT.
This risk-informed, RAT-aware scheduling framework enables the ISAC system to dynamically adapt to environmental complexity, user behavior, and radio quality, while satisfying communication and sensing performance jointly.

\section{Problem Formulation}
In this section, we formulate a joint optimization problem for vision-aided \isacname{} systems deployed in \systemname{}. 
The objective is to maximize system utility by jointly optimizing EE and perception efficiency (PE), while satisfying heterogeneous quality-of-service (QoS) constraints across agents.

\subsection{System State Representation}
The global system state at each decision epoch $t$ can be defined as ${\mathcal S}_t=\{\mathbf{z}_{\text{vis},t}^n, {\theta}_{\text{vis},t}^n, {\mathbf H}_t^n, {\mathbf H}_{\text{vis},t}^n, \gamma_t^n, \psi_t^n, d_t^n, v_t^n\}_{n\in{\mathcal N}}$, where ${\mathbf H}_t^n$ represents channel response (LTE and mmWave) and $\gamma_t^n$ is the received SINR under current RAT. 
${\mathbf H}_{\text{vis},t}^n$ is the risk-aware visual heatmap, where each spatial cell reflects the level of environmental dynamics, potential obstruction, or agent interaction complexity.
The risk heatmap provides a global view of scene dynamics and serves as a scheduling prior. Two key mechanisms are adopted:
\begin{itemize}
\item[--] \textbf{Channel Prioritization:} Agents located in high-risk zones are assigned more stable or robust communication links (e.g., mmWave) to ensure service continuity.
\item[--] \textbf{Access Reconfiguration:} Agents with low visual confidence, such as those under severe occlusion are proactively rescheduled, either by switching to alternative RATs (e.g., LTE).
\end{itemize}



\subsection{Problem Formulation}
The primary objective is to optimize both RAT selection and resource allocation with a focus on semantic profile aware EE and PE.

Follow in \cite{venturino2014energy, gao2019dynamic}, the global EE, defined as the ratio between the network sum-rate and the network power consumption, i.e.,
\begin{align}\label{eq:13}
    \text{EE}=\frac{\sum_{n=1}^NR_n}{\sum_{n=1}^Np_{n}B_n}.
\end{align}
Likewise, we can define PE for ranging and speed measurement as follows:
\begin{align}\label{eq:14}
    \mathrm{PE}^{d}(\gamma)=\sum_{n=1}^N\frac{R_n(\gamma_n)}{\kappa+\text{CRB}_d(\gamma_n)},
\end{align}
\begin{align}\label{eq:15}
    \mathrm{PE}^{v}(\gamma)=\sum_{n=1}^N\frac{R_n(\gamma_n)}{\kappa+\text{CRB}_v(\gamma_n)},
\end{align}
where $\text{CRB}(\gamma)$ represents the parameter estimate Cramero bound when the SNR is $\gamma$.
$\kappa$ is a preset constant to limit the maximum value of $\mathrm{PE}^{d}$ and $\mathrm{PE}^{v}$.
The expressions of traversing CRB for ranging and speed measurement based on pilot signals can be shown as
\begin{align}\label{eq:16-1}
    \mathrm{CRB}_{d}(\gamma_n)=&\frac{c^{2} \exp \left(-A_{s} /\left(2 \sigma_{2}^{2}\right)\right) {_1}F_1\left[1 / 2 ; 1 ; A_{s} /\left(2 \sigma_{2}^{2}\right)\right]}{8 \sqrt{2 \sigma_{2}^{2}} \pi^{3 / 2} \gamma_{n} s_{\mathrm{res}} B_{\mathrm{rms}}^{2}} \\ \notag
    &\cdot \frac{1}{L_{p}},
\end{align}
\begin{align}\label{eq:17}
    \mathrm{CRB}_{v}(\gamma_n) = &\frac{6\lambda^2\left(-A_{s} /\left(2 \sigma_{2}^{2}\right)\right) {_1}F_1\left[1 / 2 ; 1 ; A_{s} /\left(2 \sigma_{2}^{2}\right)\right]}{32\sqrt{2\sigma_2^2}\pi^{3 / 2} \gamma_{n}s_{\mathrm{res}}T_s^2}\cdot\\ \notag
    &\frac{1}{L_p(L_p+1)(2L_p+1)},
\end{align}
where $c$ is the light speed, $\lambda$ is the wavelength of carrier wave, $s_{rcs}$ denotes the radar cross-sectional area.
$T_s$ represents the symbol time interval, $B_{rms}$ is the rms bandwidth, $A_s$ represents the main path channel strength between the UAV and detected object.
$\sigma_2^2$ is the variance of the perception channel and $_1F_1(\cdot)$ presents confluent hypergeometric function.

By incorporating these PE metrics into our optimization problem, we aim to balance EE with the quality of perceptions, ensuring that the network not only operates efficiently but also meets the performance expectations of the users in the \systemname{}.
\begin{align}
    &\max_{\mathbf{w}}\lambda^{EE}\text{EE}+\lambda^{PE}(\mathrm{PE}^{d}+\mathrm{PE}^{v})\label{eq:18}\\
    \text{s.t.~}&R_n \ge R_{n,t}^{\min},\forall n \in \mathcal{N}\\
    &p_{n}\le p^{\max},\forall n \in \mathcal{N} \\
    &\sum_{n=1}^N x_n \leq N_{\text{mm}}^{\max}, \quad \sum_{n=1}^N (1 - x_n) \leq N_{\text{LTE}}^{\max},\label{eq:18-1}
\end{align}
where $\lambda^{EE},\lambda^{PE}\in [0,1]$ are balance parameter which satisfies $\lambda^{EE}+\lambda^{PE}=1$.
where $p^{\max}$ and $R_{n,t}^{\min}$ represent the maximum allowable transmit power of the UAV and the minimum data rate required for transmission based on the agent's behavior at epoch $t$, respectively.

The formulated optimization problem in \eqref{eq:18}-\eqref{eq:18-1} is inherently non-convex and challenging to solve due to several reasons. 
First, the presence of binary RAT selection variables $x_n \in \{0,1\}$ introduces combinatorial complexity, rendering the feasible solution space exponentially large with the number of agents. 
Second, the data rate $R_n$, which depends nonlinearly on SINR, is entangled with both the beamforming vectors $\mathbf{w}_n$ for mmWave agents and the transmit powers $p_n$ for LTE agents, leading to a tightly coupled and non-convex optimization landscape. 
Third, the PE terms $\mathrm{PE}^{d}$ and $\mathrm{PE}^{v}$ involve inverse CRBs, which are themselves nonlinear functions of SINR, further complicating the utility landscape. 
Lastly, the joint consideration of communication efficiency and sensing accuracy imposes a trade-off between throughput maximization and radar observability, making traditional convex optimization methods inapplicable. These challenges necessitate a scalable and adaptive optimization strategy, which we address in the next section via a learning-based approach.




\section{Algorithm Design: Vision-Aided Cross-Modal Resource Control}

To solve the non-convex joint optimization problem \eqref{eq:18}-\eqref{eq:18-1}, we propose a vision-aided learning-based algorithm that integrates high-level visual semantics, radar feedback, and communication feedback for risk-aware access control and power allocation in UAV-assisted ISAC networks. 
Our method adopts a two-stage decision pipeline, comprising: (i) a visual-semantic reconstructor for risk-aware scene profiling, and (ii) a multi-objective actor-critic learning agent for access control under energy and perception constraints.
\begin{figure*}[!t]
\centering
\includegraphics[width=0.9\textwidth]{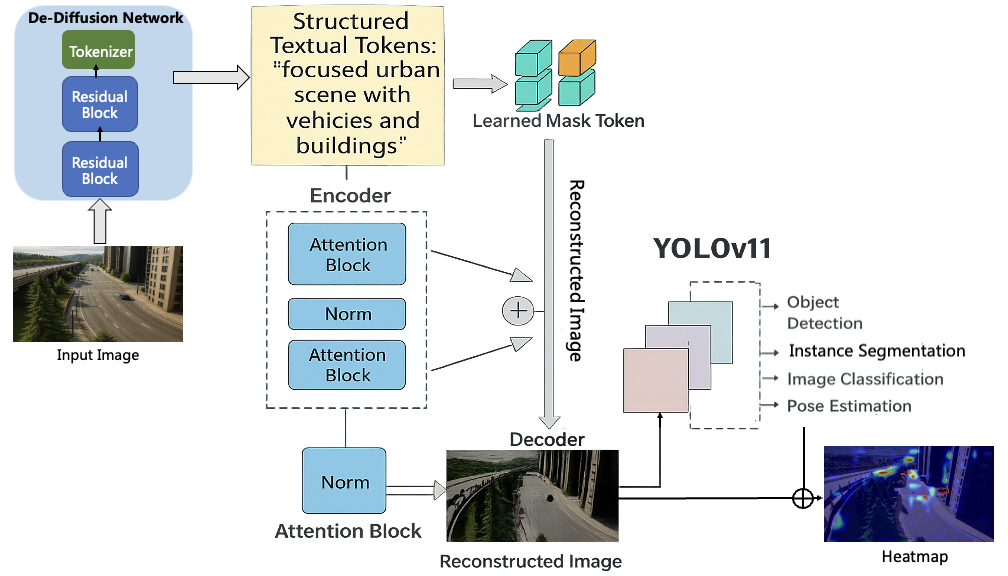}
\caption{
End-to-end pipeline for generating semantic risk-aware heatmaps from onboard visual inputs. 
The UAV first captures an input image, which is processed by a de-diffusion network to extract structured textual tokens representing privacy-preserving scene semantics (e.g., “urban scene with vehicles and buildings”). 
These tokens are uploaded to the cloud and reconstructed into synthetic imagery via a text-to-image module composed of a transformer-based encoder-decoder architecture. 
The reconstructed image is then fed into a pretrained YOLOv11 model for downstream tasks including object detection, instance segmentation, image classification, and pose estimation. 
The outputs are further fused to generate a spatial heatmap reflecting agent density, motion activity, and occlusion level, which serves as a prior for risk-aware access and resource allocation.
}
\label{fig:heat}
\end{figure*}
\subsection{Stage I: Visual-Semantic Reconstruction for Risk Map Formation}

At each scheduling epoch, the UAV captures onboard images of the operating area.
Rather than uploading raw images, we apply a masked de-diffusion model to extract privacy-preserving tokens  $\mathbf{z}_{\text{text}}^n$, which encode agent-level semantics such as heading orientation $\theta_{\text{vis}}^n$, semantic type $\text{sem}_n$, and current activity class $\text{act}_n$.
The textual tokens are then uploaded to the cloud for reconstruction into synthetic imagery $\hat{I}_t^n$ via a pretrained text-to-image diffusion model (e.g., StableXL). 
The mathematical formulation of the diffusion sampling procedure is detailed in Appendix~A. 
Subsequently, we apply a $\bf{YOLOv11}\mbox{-based}$ semantic parser on $\hat{I}_t^n$ to detect
\begin{itemize}
\item[$\dagger$] agent type (e.g., pedestrian, cyclist) ,
\item[$\dagger$] local density and occlusion,
\item[$\dagger$] crowding behavior and bounding box overlap.
\end{itemize}

These features are fused with mmWave radar outputs to construct a spatio-temporal heatmap $\mathbf{H}_{\text{vis}}(x, y)$, representing motion complexity, visual uncertainty, and potential NLoS risk. 
This heatmap guides prioritization during learning.
An example of end-to-end pipeline for generating semantic risk-aware heatmaps from onboard visual inputs is shown in Fig. \ref{fig:heat}.
\subsection{Stage II: DDPG-Based Risk-Aware Access Optimization}

We formulate the cross-modal resource scheduling problem as a Markov Decision Process (MDP), and adopt a DDPG algorithm to learn adaptive control policies. 
The decision process components are defined as follows:

{\bf{\em State Space.}} At each time step $t$, the global state vector $s_t$ aggregates cross-modal observations for all agents:
\begin{align}
s_t = \left\{ \mathbf{z}_{\text{vis},t}^n, \theta_{\text{vis},t}^n, \gamma_t^n, d_t^n, v_t^n, \psi_t^n \right\}_{n \in \mathcal{N}},
\end{align}
where $\mathbf{z}_{\text{vis},t}^n$ and $\theta_{\text{vis},t}^n$ denote the semantic profile and heading orientation, respectively,  $\gamma_t^n$ the SINR, $d_t^n$ and $v_t^n$ are radar-estimated distance and Doppler velocity, and $\psi_t^n$ the angle from the radar.

{\bf{\em Action Space.}} The action vector $\mathbf{a}_t$ for all agents includes:
\begin{align}
\mathbf{a}_t = \left\{ x_t^n, p_t^n, \mathbf{w}_t^n \right\}_{n \in \mathcal{N}},
\end{align}
where $x_t^n \in \{0,1\}$ is the RAT assignment (1 for mmWave, 0 for LTE), $p_t^n$ is the transmit power, and $\mathbf{w}_t^n$ is the beamforming vector selected from codebook $\mathcal{B}$ using visual heading priors.

{\bf{\em Reward Function.}} To jointly account for communication quality, sensing accuracy, and semantic consistency, we define the following multi-objective reward for each agent $n$ at time step $t$:
\begin{align}\label{eq:28}
r_t^n = \lambda_{\text{EE}} \cdot \text{EE}_t^n + \lambda_{\text{PE}} \cdot \text{PE}_t^n + \lambda_{\text{SR}} \cdot \text{SR}_t^n,
\end{align}
where $\text{SR}_t^n = \mathbb{I}(\gamma_t^n > \gamma_{\text{th}})$ serves as a link stability indicator, capturing whether agent $n$ maintains reliable SINR. 
Agents located in high-risk regions identified through the semantic risk heatmap (e.g., motion-intensive or visually occluded zones) are prioritized for allocation to more stable communication links (e.g., LTE or fallback mmWave beams). 
The risk heatmap is constructed from visual semantic cues and used as a scheduling prior to guide access control decisions. 
The balance parameters $\lambda_{\text{EE}}, \lambda_{\text{PE}}, \lambda_{\text{SR}} \in [0,1]$ satisfy $\lambda_{\text{EE}} + \lambda_{\text{PE}} + \lambda_{\text{SR}} = 1$ and are used to adjust the emphasis of each component. Constraint violations (e.g., $R_t^n < R_{n,t}^{\min}$) incur a heavy penalty $r_t^n = -100$.

\subsection{Training and Deployment Protocol}

The proposed DDPG framework is trained using an off-policy actor-critic strategy with experience replay and soft target updates. During training, the edge cloud interacts with a simulated LAENet environment and collects a sequence of transitions $\left(s_t, a_t, r_t, s_{t+1}\right)$, which are stored in a replay buffer. At each training iteration, a mini-batch of transitions is sampled from the buffer for gradient-based updates.

For each transition in the batch, the target value for the critic is computed as:
\begin{align}
y_t = r_t + \gamma Q'\left(s_{t+1}, \mu'(s_{t+1}; \theta_{\mu'}) ; \theta_{Q'} \right),
\end{align}
where $Q'$ and $\mu'$ are the target critic and target actor networks, $\gamma$ is the temporal discount factor, and $r_t$ is the multi-objective reward defined in Eq. \eqref{eq:28}. The critic loss function is defined as the mean-squared temporal difference (TD) error:
\begin{align}
\mathcal{L}_Q = \frac{1}{B} \sum_{i=1}^B \left( Q(s_t^i, a_t^i; \theta_Q) - y_t^i \right)^2,
\end{align}
where $B$ is the batch size. The critic network parameters $\theta_Q$ are updated via gradient descent to minimize $\mathcal{L}_Q$.

The actor is updated using the sampled policy gradient, which maximizes the expected return under the current critic evaluation:
\begin{align}
\nabla_{\theta_\mu} J \approx \frac{1}{B} \sum_{i=1}^B \nabla_a Q(s, a; \theta_Q) \big|_{s = s_t^i, a = \mu(s_t^i)} \cdot \nabla_{\theta_\mu} \mu(s_t^i; \theta_\mu).
\end{align}
To stabilize training, the target networks are updated using a soft-update mechanism:
\begin{align}
\theta_{Q'} \leftarrow \tau \theta_Q + (1 - \tau) \theta_{Q'}, \quad \theta_{\mu'} \leftarrow \tau \theta_\mu + (1 - \tau) \theta_{\mu'},
\end{align}
where $\tau \ll 1$ is the target update rate.

After convergence, the trained policy $\mu(s)$ is deployed onboard the edge controller. 
During online execution, the agent receives real-time state observations, including updated visual tokens and mmWave radar estimates, and directly infers the optimal resource allocation action $\mathbf{a}_t$ without requiring further gradient updates. 
The complete vision-aided risk-aware resource optimization algorithm is detailed in Algorithm \ref{algorithm:1}. 

The proposed algorithm introduces several key innovations that distinguish it from conventional cross-layer designs. First, it integrates de-diffused visual semantics extracted through privacy-preserving masked generation with mmWave radar observations to construct rich cross-modal state representations that drive both access and scheduling decisions. 
Second, it defines a novel semantic risk-aware reward function that jointly accounts for communication efficiency, radar-based perception quality, and cross-modal reliability by penalizing inconsistent heading estimates and uncertain visual contexts. 
Finally, the algorithm employs a deterministic policy gradient framework to enable fine-grained, continuous control over beamforming vectors and power allocation, making it well-suited for highly dynamic and infrastructure-sparse LAE environments.
\begin{algorithm}[!t]
\SetAlgoLined
\caption{De-Diffusion-Driven Vision-Aided Risk-Aware Resource Optimization Algorithm (DeDiff-VARARO)}
\label{algorithm:1}
\KwIn{
UAV’s maximum power constraint $p^{\max}$;\\

Pretrained de-diffusion and diffusion models;\\

YOLOv11 semantic parser and SlowFast activity classifier;\\

Pretrained actor-critic networks $(\mu, Q)$ for DDPG.
}
\While{Agent $n$ is detectable}{
    Estimate $(d_t^n, v_t^n, \psi_t^n)$ via mmWave radar.\\
    Capture raw image $I_t^n$ from onboard camera.\\
    Extract semantic token $\mathbf{z}_{\text{text}}^n$ via masked de-diffusion.\\
    Upload token to server and reconstruct image $\hat{I}_t^n$.\\
    \begin{itemize}
  \item  Apply {\bf YOLOv11} to detect agent type,\\
   \item Apply {\bf SlowFast} to classify agent activity.\\
    \end{itemize}
    Construct structured semantic profile $\mathrm{sf}_n = (\text{sem}_n, \text{act}_n)$.\\
    Generate risk-aware heatmap $\mathbf{H}_{\text{vis}}(x, y)$ based on visual features.\\
    Fuse $\text{sem}_n$, radar data, and SINR $\gamma_t^n$ into state vector $s_t^n$.\\
    Compute semantic reliability: $\text{SR}_t^n = \mathbb{I}(\gamma_t^n > \gamma_{\text{th}})$.\\
    Evaluate reward: 
    \begin{align}
    r_t^n = \lambda_{\text{EE}} \cdot \text{EE}_t^n + \lambda_{\text{PE}} \cdot \text{PE}_t^n + \lambda_{\text{SR}} \cdot \text{SR}_t^n.\notag
    \end{align}
    Use actor network to generate action:
    \begin{align}
    a_t^n = \mu(s_t^n): \quad x_t^n\text{ (RAT)},\ p_t^n\text{ (power)},\ \mathbf{w}_t^n\text{ (beam)}.\notag
    \end{align}
    Allocate $\mathbf{w}_t^n$ to antenna array.\\
    Store $(s_t^n, a_t^n, r_t^n, s_{t+1}^n)$ into replay buffer for training.\\
}
\KwOut{
Optimal RAT selection, power allocation, and beamforming configuration.
}
\end{algorithm}

\section{Simulation Results}
\subsection{Simulation Environments and Settings}
To evaluate the performance of the proposed vision-aided ISAC framework, we simulate a dynamic \systemname{} populated with multiple mobile agents of varying semantic types and activities.

{\bf\em{Semantic Dataset.}}  To construct comprehensive semantic tokens for each detected agent, we leverage a combination of publicly available datasets and pretrained detection models. 
Specifically, we adopt the MS-COCO dataset \cite{coco} to define an \emph{agent type set} comprising common mobile entities such as $\{\text{pedestrian}, \text{bicycle}, \text{motorcycle}, \text{car}, \text{bus}\}$. 
A {\bf YOLOv11}-based detector, pretrained on MS-COCO and fine-tuned on urban scenes, is applied to each reconstructed image $\hat{I}_t^n$ to infer the agent type label.
In parallel, we utilize the AVA v2.2 dataset  \cite{gu2018ava} to define an \emph{activity set} containing over 60 atomic human actions, including examples such as $\{\text{standing}, \text{walking}, \text{talking}, \text{running}, \text{carrying}\}$. 
A {\bf SlowFast}-based action recognition model is employed to classify the most probable activity within each bounding box. 
The final semantic profile for agent $n$ is constructed as a structured tuple:
\begin{align}
\mathbf{z}_{\text{vis}}^n = \left(\text{sem}_n, \text{act}_n\right),
\end{align}
which encapsulates both the physical agent category and its observed behavior. 
These tokens are subsequently embedded via a text encoder into a unified latent representation $\mathbf{z}_{\text{text}}^n$ for downstream risk-aware resource allocation.

{\bf\em{mmWave Radar and Communication Parameters.}}  The wavelength $\lambda$ is set to $2$mm while the number of pilot sumbols $L=14$.
The symbol time interval $T_s$ is set to $0.05$ms, and the radar cross-sectional area $s_{rcs}$ is $100\text{m}^2$. 
The variance of communication channel $\sigma_1^2=2$, and the Rice factor of perception channel $K=A_s/\sigma_2^2=3$ while the rms bandwith $B_{rms}=\sqrt{12}B_n$.

{\bf\em{Learning Framework.}}  The Actor network consists of two fully connected layers, where it processes the input state and outputs an action using Rectified Linear Unit (ReLU) and Sigmoid activation functions respectively, with the Sigmoid ensuring that the action values are within a specified range.
On the other hand, the Critic network, also comprising two fully connected layers, takes both the state and action as inputs, merges them, and then outputs a single value representing the estimated value of the state-action pair, using a ReLU activation function in its first layer.
The learning rate is set to $0.001$, discount factor is set to $0.99$ and the variance of explore noise is $0.2$.
The soft update factor is $0.005$.
The batch size is set to $64$ and the memory buffer size is $10000$.
To clearly delineate the parameters of the DDPG algorithm, we have enumerated the hyperparameters in Table \ref{table:2}.

All simulations are conducted using PyTorch-based DDPG implementation, and training converges within $3000$ steps. 
\begin{table}[!t]
\centering
\caption{Hyperparameters used in DDPG}
\begin{tabular}{c|c|c}
  \hline
  {\bf Symbol} & {\bf Expression} & \bf{Value} \\
  \hline
  $\eta_a$ & Learning rate for actor network& $0.001$\\
  \hline
  $\eta_c$ & Learning rate for critic network& $0.001$\\
  \hline
  $\gamma$ & Discount factor & $0.99$\\
  \hline
  $\tau$ & Soft update parameter & $0.005$\\
  \hline
  $RB$ & Replay buffer size & $10000$\\
  \hline
  $BS$ & Batch size & $64$\\
  \hline
  $ES$& Max step & $3000$\\
  \hline
  $noise$ & Explore noise & $0.2$\\
  \hline
\end{tabular}
\label{table:2}
\end{table}
\begin{figure*}[!t]
\centering
\subfigure[]{
\begin{minipage}[t]{0.45\linewidth}
\includegraphics[width=3.5in]{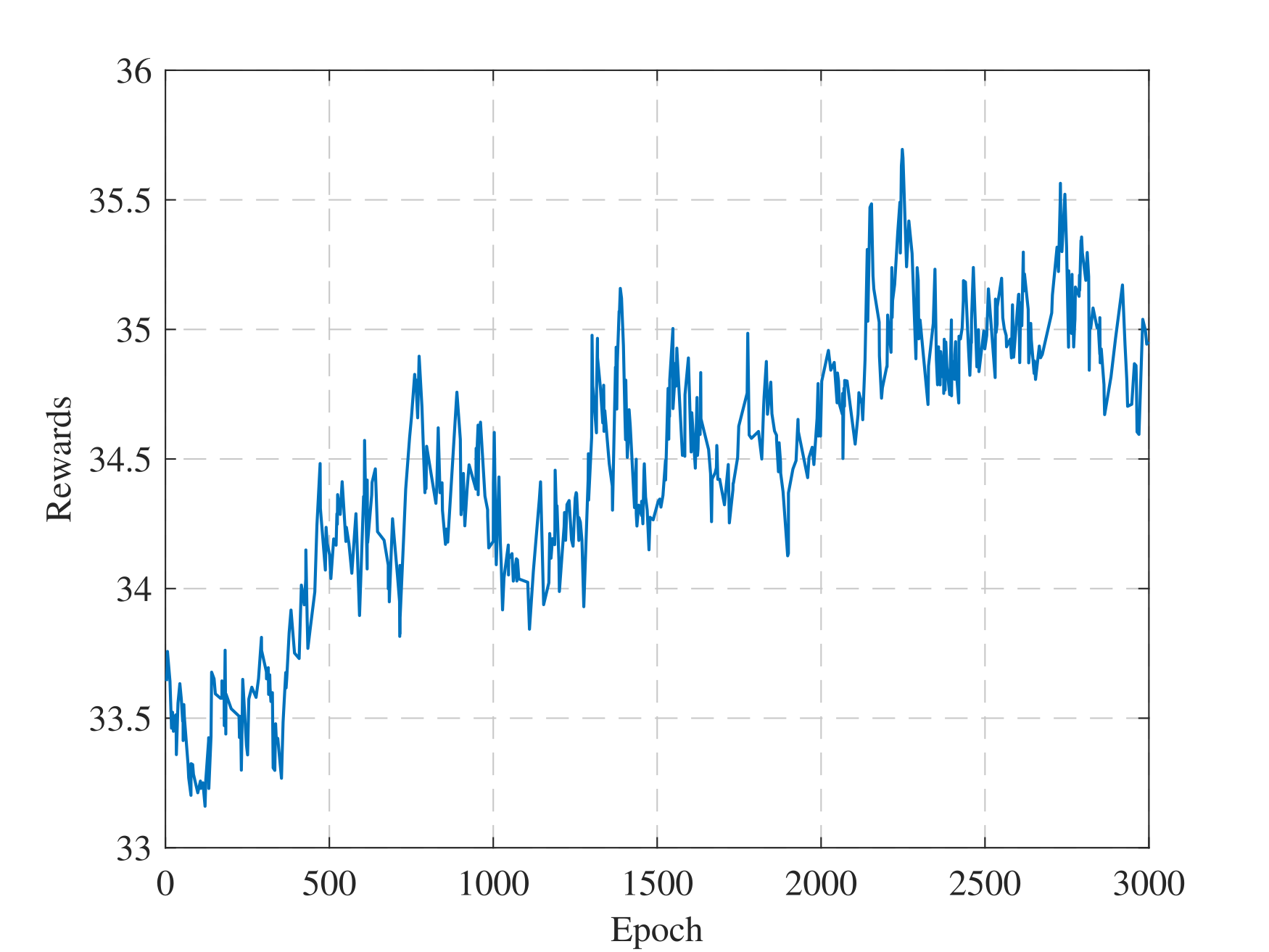}
\end{minipage}
\label{fig:3-1}
}
\subfigure[]{
\begin{minipage}[t]{0.45\linewidth}
\includegraphics[width=3.5in]{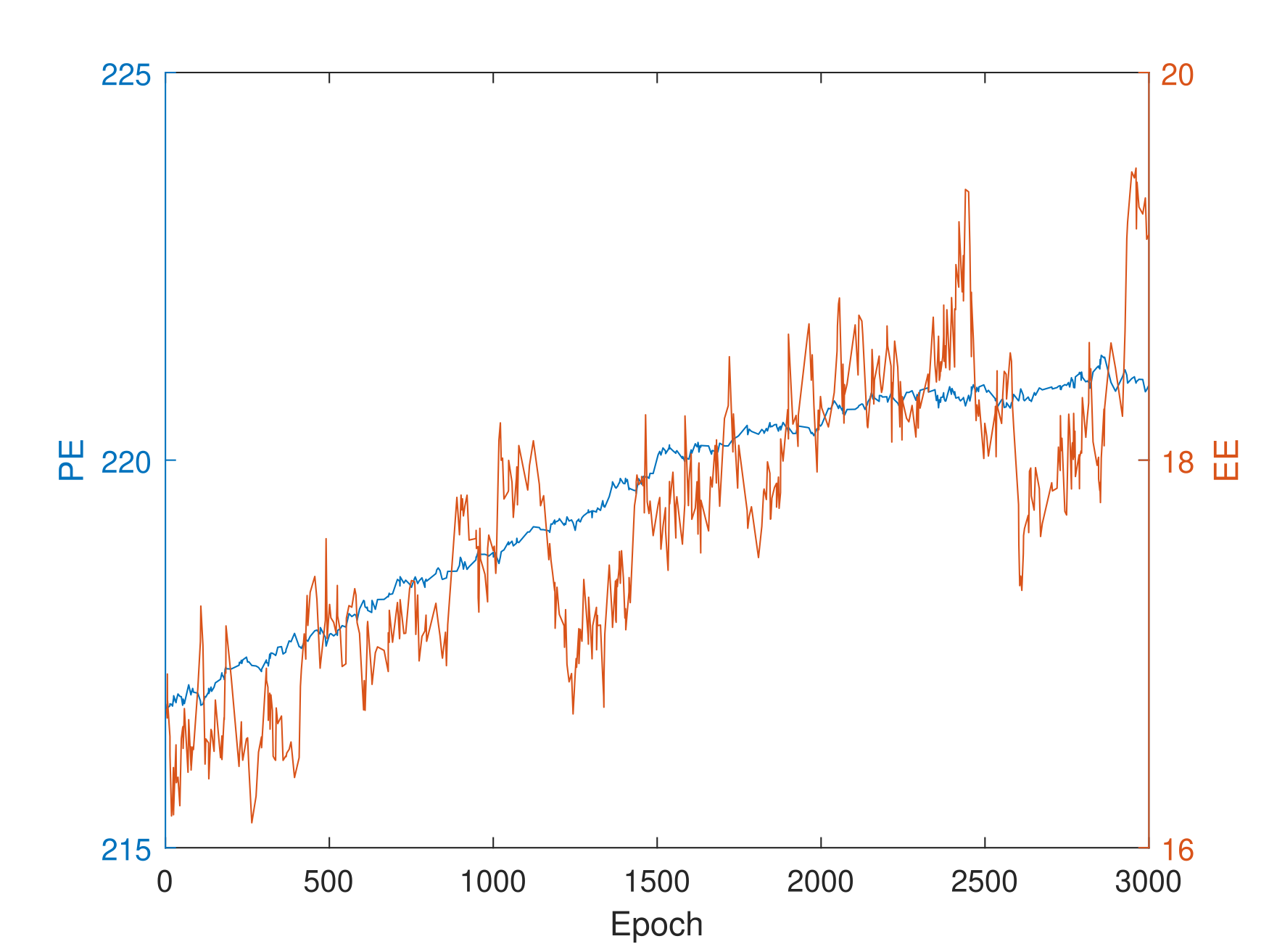}
\end{minipage}
\label{fig:3-2}
}
\caption{(a)Convergence Validation, (b)Energy efficiency and perception quality over epochs.}
\end{figure*}

\subsection{Comparison Baselines}
In order to gain insight into the performance of the proposed \methodname{}, we compare it against six relevant baseline methods using different visual-token generation strategies and semantic profile modeling mechanisms. 
For reference, we also include a raw-image-based upper bound that directly leverages full visual input without semantic compression.

\begin{itemize}
\setlength{\itemsep}{1pt}
\setlength{\parskip}{1pt}
\setlength{\parsep}{1pt}
\item {\bf\em Raw Image (Direct Vision-Based)}: this method bypasses the semantic compression pipeline and directly utilizes full-resolution visual inputs for agent profile recognition. 
The extracted features are then fed into the DDPG algorithm for resource allocation, serving as an oracle-style upper bound for performance comparison.
\item {\bf\em Random}: both the semantic profile, RAT selection, and precoding matrix are randomly determined.
\item {\bf\em Semantic Profile Ignored}: only the joint optimization of RAT selection and precoding matrix are considered where agents' semantic profiles are randomly selected.  
\item {\bf\em \methodname{} \& Copilot \cite{dakhel2023github} (respectively, \methodname{} \& StableXL \cite{zhuang2023pilot})}: In the proposed \deisacname{} framework in \systemname{}, the structured textual tokens $\mathbf{z}_{\text{text}}^n$ are extracted by De-Diffusion model and the process of text-to-image is executed by well-trained Copilot (respectively, StableXL). Subsequently, the proposed \vddpgname{} is used to semantic profile recognition, RAT selection, and precoding matrix optimization.
\item {\bf\em \vddpgname{} with ChatGPT \& Copilot (respectively, \vddpgname{} with ChatGPT\&StableXL)}: In the proposed vision-aided \isacname{} framework in \systemname{}, the semantic tokens are extracted by ChatGPT and the process of text-to-image is executed by well trained Copilot (respectively, StableXL). Subsequently, the proposed \vddpgname{} is used to semantic profile recognition, RAT selection, and precoding matrix optimization.
\end{itemize}
\begin{figure}[!t]
\centering
\includegraphics[width=0.5\textwidth]{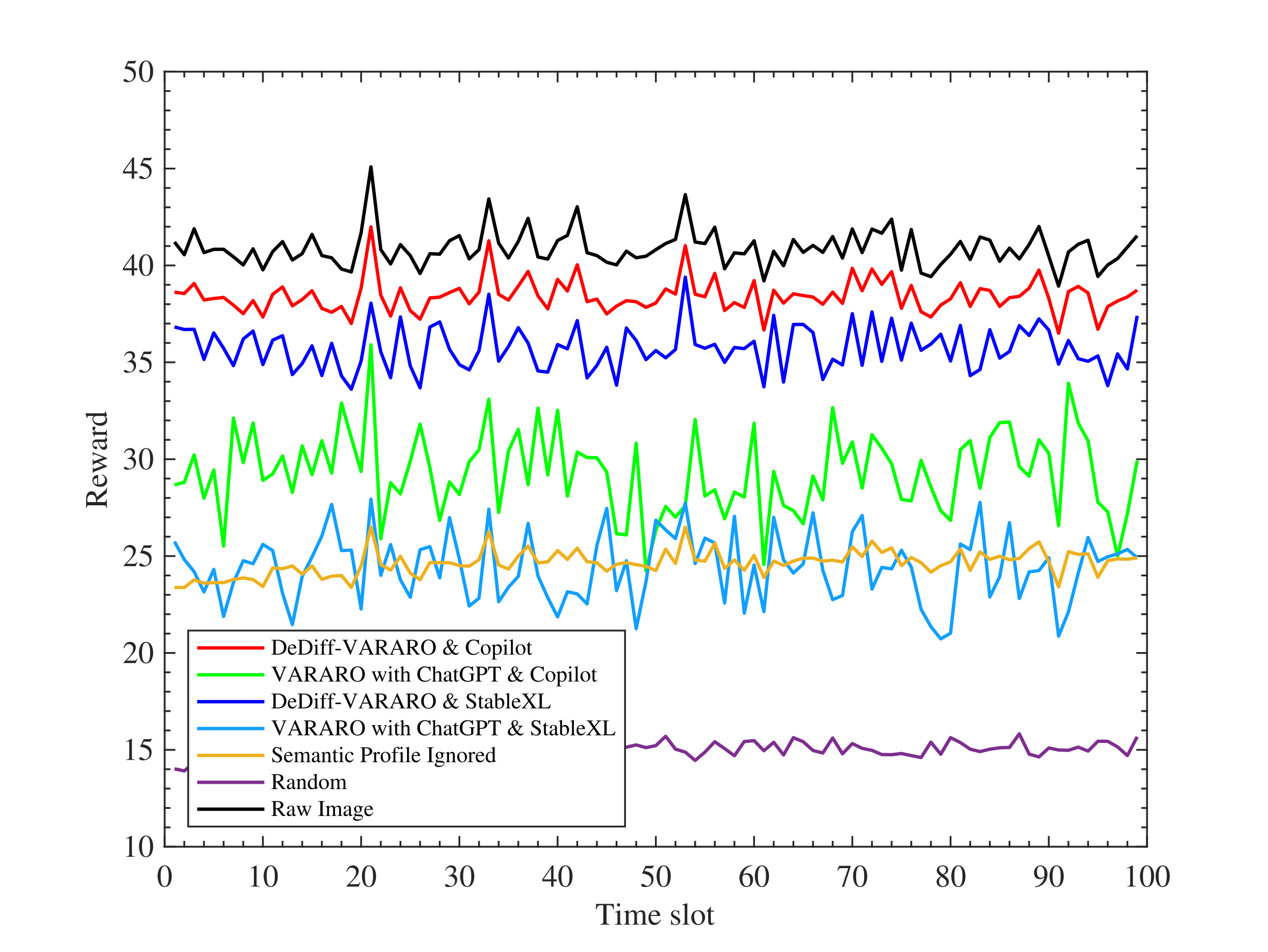}
\caption{Comparison of average reward over 100 time slots under different semantic generation and access control strategies.}
\label{fig:3-3}
\end{figure}
\subsection{Results and Discussion}
\subsubsection{Effectiveness of Proposed \methodname{} Algorithm}
The test reward curve of the proposed \methodname{} in \systemname{} is presented in Fig. \ref{fig:3-1}. 
It is clear from Fig. \ref{fig:3-1} that the semantic profile and risk heat map embedded in \vddpgname{} is effectively learning and refining its policy to enhance the RAT selection, power allocation, and precoding matrix optimization. 
Fluctuations during the training process indicates an active exploration strategy, which gradually stabilizes, showing that the \methodname{} is converging towards a consistent and efficient policy. 
The trend demonstrates the algorithm's ability to balance EE and PE. 
Fig. \ref{fig:3-2} shows the performance of EE and PE of \methodname{} under vision-aided \isacname{} framework in \systemname{} versus the training epoch. 
From the results, we observe that despite inherent trade-offs, the algorithm is making effective progress towards the simultaneous optimization of both objectives (EE and PE). 

\subsubsection{Performance Demonstration Versus Time Slot}
Fig.~\ref{fig:3-3} illustrates the time-averaged reward performance of the proposed \methodname{} algorithm under different semantic generation strategies, in comparison with several baseline methods. 
The two variants of our method, DeDiff-VARARO with Copilot and DeDiff-VARARO with StableXL, demonstrate consistently superior performance across all time slots, benefiting from the privacy-preserving semantic tokens and agent semantic profile-aware decision policies. 
Notably, the Copilot-based version achieves the highest overall reward, indicating its advantage in generating coherent and task-relevant visual tokens.
In contrast, the baseline methods that omit de-diffusion or rely solely on ChatGPT-style tokenization without visual masking (e.g., VARARO with ChatGPT \& Copilot or ChatGPT \& StableXL) show a noticeable performance gap. These methods still capture high-level intent but lack the robustness offered by visual token regularization and semantic profile precision.
The ``Semantic Profile Ignored'' baseline, which removes semantic type and activity differentiation, performs significantly lower, underscoring the importance of structured semantic information in the control loop. 
As expected, the ``Random'' method yields the lowest reward due to its lack of adaptive scheduling, while the ``Raw Image'' configuration serves as an oracle-style reference, where full visual data is directly exploited without compression or privacy filtering. 

The \methodname{} \& Copilot method yields an average reward within $4\%$ of the Raw Image baseline, as measured by the relative gap metric defined in
\begin{align}
\mathrm{Gap} = \frac{\bar{r}_{\text{raw}} - \bar{r}_{\text{ours}}}{\bar{r}_{\text{raw}}} \times 100\%,
\end{align}
where $\bar{r}_{\text{raw}}$ and $\bar{r}_{\text{dediff}}$ denote the time-averaged rewards of the Raw Image method and the proposed approach, respectively. 
This confirms that the proposed pipeline can closely approach the oracle upper bound, despite relying only on privacy-preserving semantic tokens instead of full-resolution visual input.
\begin{figure*}[!t]
\centering
\subfigure[]{
\begin{minipage}[t]{0.3\linewidth}
\includegraphics[width=2.5in]{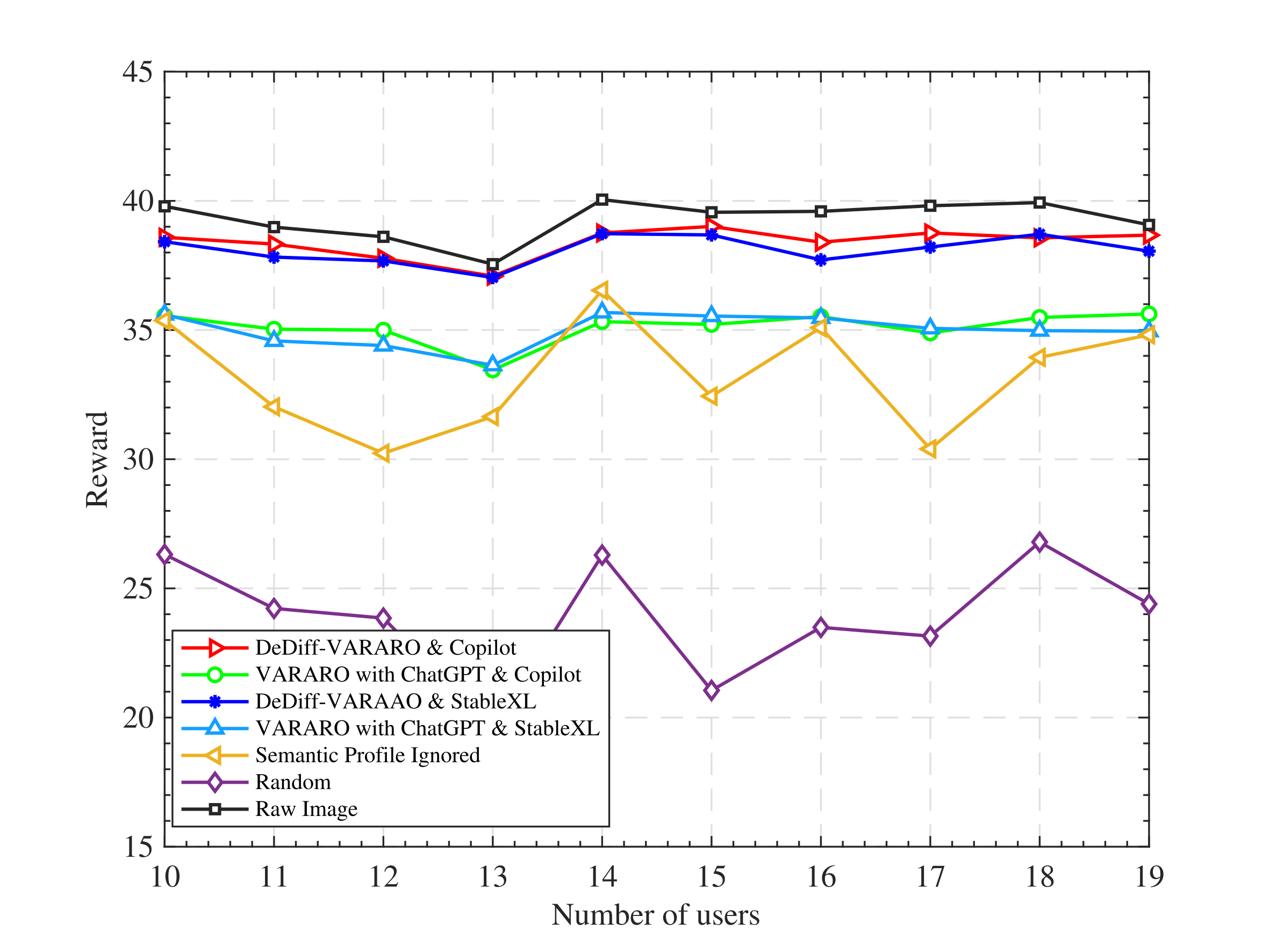}
\end{minipage}
\label{fig:4-1}
}
\subfigure[]{
\begin{minipage}[t]{0.3\linewidth}
\includegraphics[width=2.5in]{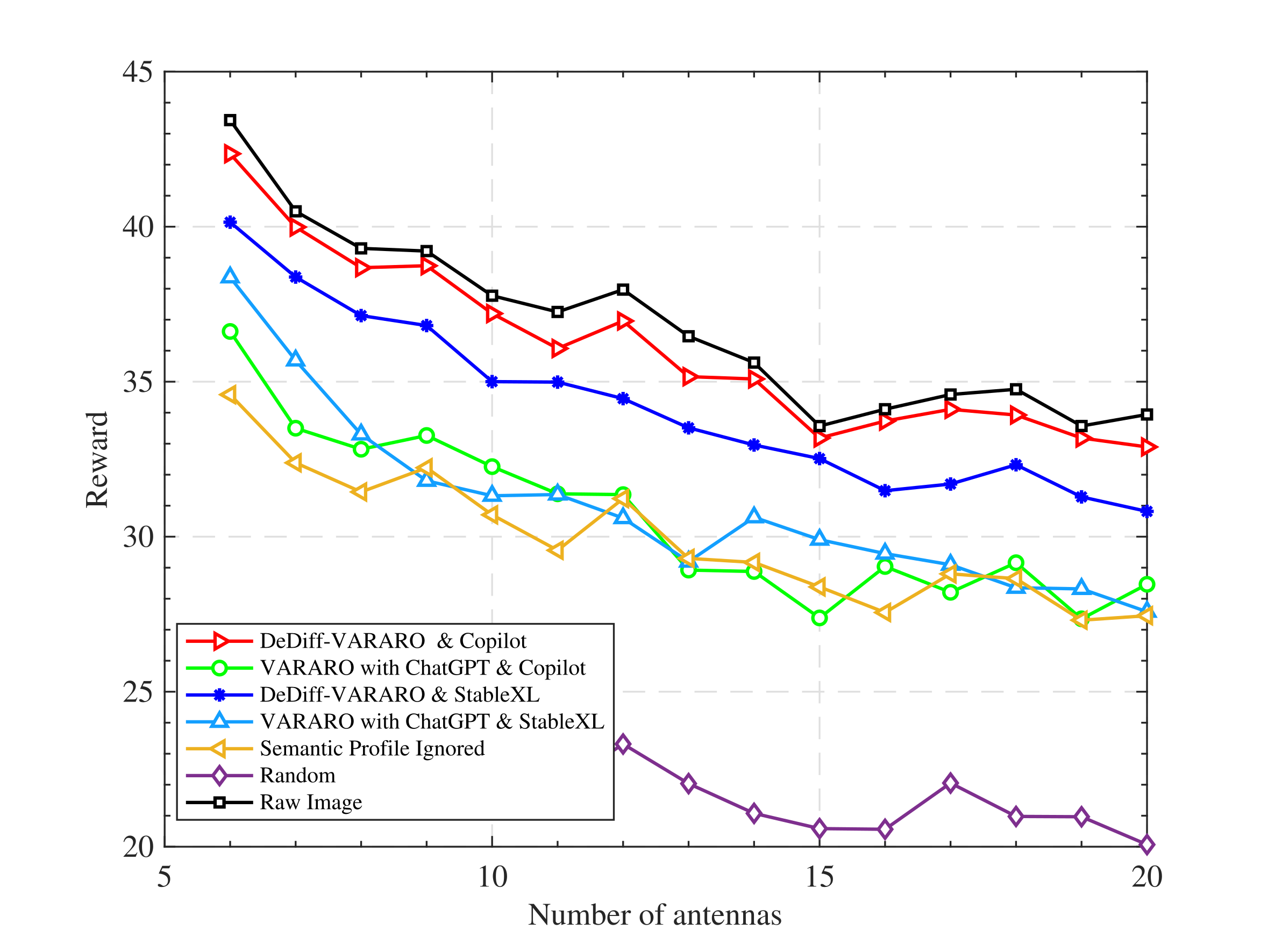}
\end{minipage}
\label{fig:4-2}
}
\subfigure[]{
\begin{minipage}[t]{0.3\linewidth}
\includegraphics[width=2.5in]{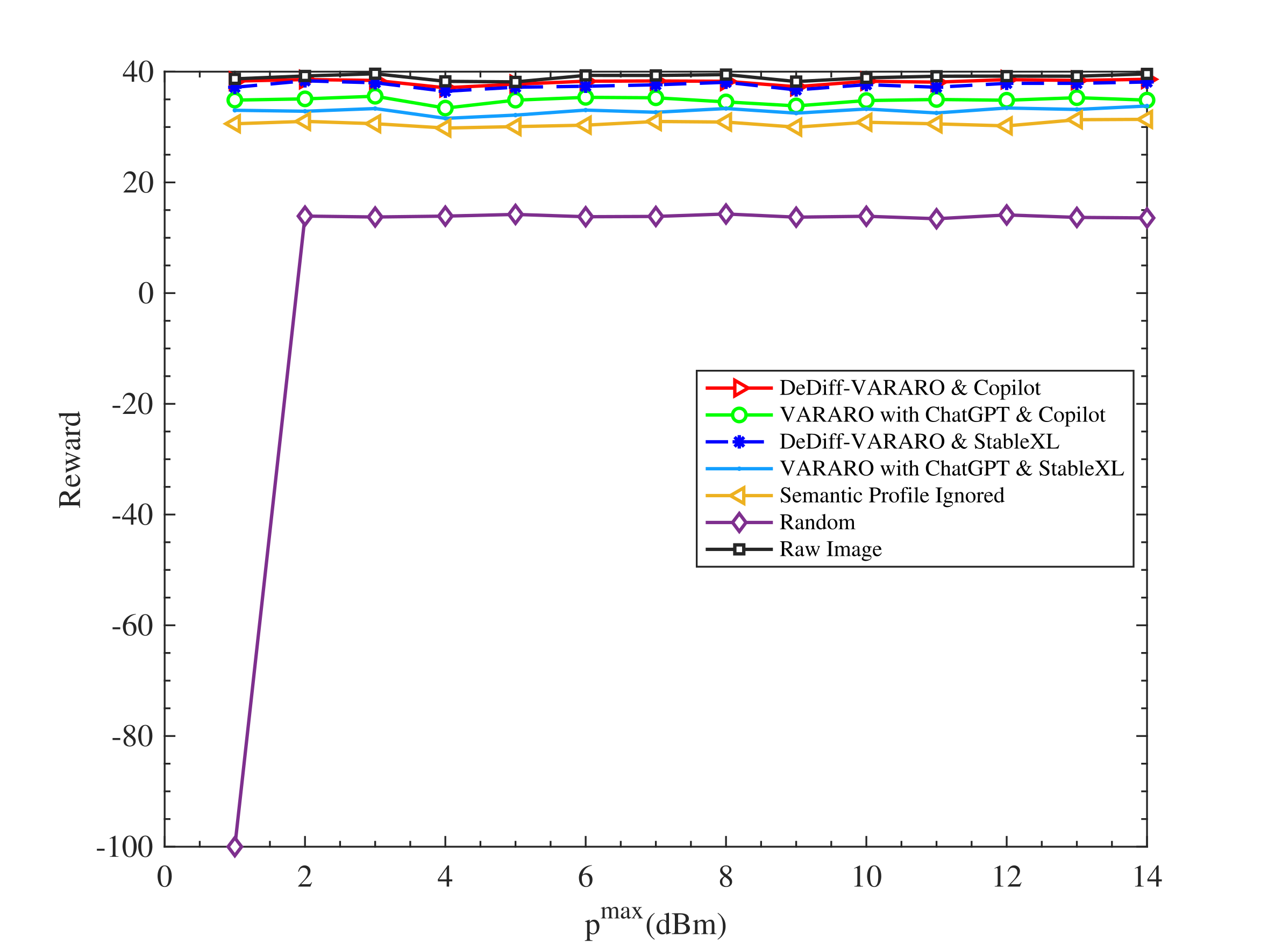}
\end{minipage}
\label{fig:4-3}
}
\caption{(a)Average reward under different $N$, (b)Average reward under different $L$, and (c) Average reward under different $p^{max}$.}
\end{figure*}

\subsubsection{Performance versus Number of agents $N$}
Fig.~\ref{fig:4-1} illustrates the reward performance as the number of agents $N$ increases from $10$ to $19$. 
The proposed \methodname{} methods maintain consistently high reward levels, exhibiting strong scalability and robustness under growing agent density. 
In contrast, the performance of the Semantic Profile Ignored and Random baselines degrades or stagnates as $N$ increases, indicating limited capacity in adapting to multi-agent interference and resource contention. 
The advantage of the proposed methods stems from their ability to allocate resources based on individualized semantic profiles, enabling fine-grained scheduling under user heterogeneity.

\subsubsection{Performance versus Number of Antennas $M$}
Fig.~\ref{fig:4-2} presents the reward variation as the number of antennas $M$ increases. 
While all methods show a general reward decline with larger antenna arrays possibly due to increased beam alignment complexity, the \methodname{}-driven methods consistently outperform the baselines across the entire range. 
Notably, the Random and Semantic Profile Ignored methods struggle to leverage spatial degrees of freedom effectively, leading to accelerated performance degradation. 
These results highlight the importance of semantic-guided beam selection in fully utilizing spatial multiplexing gains under practical constraints.

\subsubsection{Performance versus Maximum Transmit Power $p^{\max}$}
Fig.~\ref{fig:4-3} depicts the impact of transmit power constraint $p^{\max}$ on reward performance. 
The proposed methods demonstrate stable and near-optimal reward levels across the entire power range, reflecting their robustness to power scaling. 
In contrast, the Random baseline exhibits a severe reward drop when $p^{\max}$ is below $4\mbox{dBm}$, indicating an inability to handle low-power constraints. 
Above this threshold, its performance saturates. The DeDiff-VARARO’s consistent performance affirms its ability to adapt transmission decisions to power limits while preserving semantic awareness and energy efficiency.

\section{Conclusion}
In this work, we proposed a vision-aided \isacname{} framework for UAV-assisted \systemname{} that integrates semantic-level perception and cross-modal resource control. 
A masked de-diffusion model was introduced to extract privacy-preserving visual tokens encoding agent types, orientations, and activity classes, which were fused with mmWave radar feedback to construct a semantic risk heatmap for scheduling. To address dynamic access and power allocation, we formulated a multi-objective optimization problem and developed a \methodname{}-based control algorithm leveraging cross-modal states. 
Simulation results demonstrate that the proposed \methodname{}-based approach achieves near-optimal performance with strong robustness to agent density, antenna variation, and power constraints, while preserving privacy and semantic fidelity. 
These results confirm the viability of semantic token-driven control in scalable and privacy-compliant vision-aided \isacname{} systems.

\bibliographystyle{ieeetr}
\bibliography{reference}

\begin{appendices}
\section{Diffusion Process}
We denote the original data as $\boldsymbol{x}_0$, which satisfies the distribution $\boldsymbol{x}_0 \sim q(\boldsymbol{x}_0)$.
The forward diffusion process is defined by adding a small Gaussian noise to the sample at each step $t$.
The whole process is a first-order Markov process, and $\boldsymbol{x}_t$ is only related to $\boldsymbol{x}_{t-1}$, which can be expressed by
\begin{align}
    q(\boldsymbol{x}_t|\boldsymbol{x}_{t-1})=\mathcal{N}(\boldsymbol{x}_t;\sqrt{1-\beta_t}\boldsymbol{x}_{t-1},\beta_t \mathbf{I})
\end{align}
where $q(\boldsymbol{x}_t|\boldsymbol{x}_{t-1})$ denotes the condition probability of $\boldsymbol{x}_t$ under given $\boldsymbol{x}_{t-1}$, which follows a Gaussian distribution with mean $\sqrt{1-\beta_t}\boldsymbol{x}_{t-1}$ and variance $\beta_t \mathbf{I}$.
$\left \{\beta_t \in (0,1)\right \}_{t=1}^T$ is used to control the noise level of each step.
Further given $\boldsymbol{x}_0$, the condition probability of the entire Markov process is the combination of the conditional probabilities of each step, which can be expressed by
\begin{align}
q(\boldsymbol{x}_{1:T}|\boldsymbol{x}_0)=\prod_{t=1}^T q(\boldsymbol{x}_t|\boldsymbol{x}_{t-1})
\end{align}
We can further denote $\alpha_t=1-\beta_t$ and $\bar{\alpha}_t=\prod_{i=1}^ta_i$,then $\boldsymbol{x}_t$ can be formulated by
\begin{align}
    \boldsymbol{x}_t&=\sqrt{\alpha_t}\boldsymbol{x}_{t-1}+\sqrt{1-\alpha_t}\epsilon_{t-1}\\ \notag
    &=\sqrt{\alpha_t \alpha_{t-1}}\boldsymbol{x}_{t-2}+\sqrt{1-\alpha_t \alpha_{t-1}}\bar{\epsilon}_{t-2} \\ \notag
    &= \dots \\ \notag
    &=\sqrt{\bar{\alpha}_t}\boldsymbol{x}_0 + \sqrt{1-\bar{\alpha}_t}\bar{\epsilon}_t
\end{align}
where $\epsilon_{t-1},\epsilon_{t-2},\dots $ and $\bar{\epsilon}_{t} \sim \mathcal{N}(0,\mathbf{I})$.

The forward diffusion process gradually adds the noise to the original data.
If the process is reversed, we can restore the original data sample from the noise $\boldsymbol{x}_T \sim \mathcal{N}(0,\mathbf{I})$.
This is the basic idea of data generation based on the diffusion model, that is, every step from $\boldsymbol{x}_T$ to $\boldsymbol{x}_0$, given $\boldsymbol{x}_t$, sample $\boldsymbol{x}_{t-1}$ with the condition probability $q(\boldsymbol{x}_{t-1}|\boldsymbol{x}_t)$ until finally get $\boldsymbol{x}_0$.
However, the conditional probability $q(\boldsymbol{x}_{t-1}|\boldsymbol{x}_t)$ of the reverse diffusion process can also be considered to satisfy the Gaussian distribution when the noise increases at each step of the forward diffusion process is small.

In fact, we cannot solve the conditional probability directly, because the whole dataset is needed for direct solution.
In addition to solving directly, another method is to train a model $p_{\theta}$ to approximate the above condition probabilities, which can be expressed by
\begin{align}
    p_{\theta}(\boldsymbol{x}_{t-1}|\boldsymbol{x}_t)=\mathcal{N}(\boldsymbol{x}_{t-1};\mu_{\theta}(\boldsymbol{x}_t,t),\Sigma_{\theta}(\boldsymbol{x}_t,t))
\end{align}

\begin{align}
    p_{\theta}(\boldsymbol{x}_{0:T})=p(\boldsymbol{x}_T)\prod_{t-1}^T p_{\theta}(\boldsymbol{x}_{t-1}|\boldsymbol{x}_t)
\end{align}
For every $t=T,T-1,\dots,0$, we can predict the mean $\mu_{\theta}(\boldsymbol{x}_t,t)$ and variance $\Sigma_{\theta}(\boldsymbol{x}_t,t)$ of the Gaussian distribution based on the model $\theta$ and the input $\boldsymbol{x}_t$ and $t$.
Based on the prediction results, we can sample $\boldsymbol{x}_{t-1}$ from the distribution $p_{\theta}(\boldsymbol{x}_{t-1}|\boldsymbol{x}_t)$.
And so on until we finally get a possible value of $\boldsymbol{x}_0$.

Through the backward diffusion process, it is possible to generate a series of data from a random noise satisfying the Gaussian distribution $\mathcal{N}(0,\mathbf{I})$.
Since each prediction is sampled from a probability density function, the diversity of the generated data can be guaranteed.

Furthermore, we can transfer the prediction of mean $\mu_{\theta}(\boldsymbol{x}_t,t)$ and variance $\Sigma_{\theta}(\boldsymbol{x}_t,t)$ to the prediction of noise $\epsilon_{\theta}(\boldsymbol{x}_t,t)$ and we can derive the relationship between $\mu_{\theta}(\boldsymbol{x}_t,t)$ and $\epsilon_{\theta}(\boldsymbol{x}_t,t)$.
\begin{align}
    \mu_{\theta}(\boldsymbol{x}_t,t)=\frac{1}{\sqrt{\alpha_t}}\left( \boldsymbol{x}_t - \frac{1-\alpha_t}{\sqrt{1-\bar{\alpha}_t}}\epsilon_{\theta}(\boldsymbol{x}_t,t)\right)
\end{align}
then the $p_{\theta}(\boldsymbol{x}_{t-1},\boldsymbol{x}_{t})$ can be expressed as
\begin{align}
    &p_{\theta}(\boldsymbol{x}_{t-1},\boldsymbol{x}_{t})=\\ \notag
    &\mathcal{N}\left(\boldsymbol{x}_{t-1};\frac{1}{\sqrt{\alpha_t}}\left(\boldsymbol{x}_t-\frac{1-\alpha_t}{\sqrt{1-\bar{\alpha}_t}}\epsilon_{\theta}(\boldsymbol{x}_t,t)\right),\Sigma_{\theta}(\boldsymbol{x}_t,t)\right)
\end{align}
We can set the $\Sigma_{\theta}(\boldsymbol{x}_t,t)$ as a constant and predict $\epsilon_{\theta}(\boldsymbol{x}_t,t)$ via model.

\end{appendices}
    
\end{document}